# Depolarized Holography with Polarization-multiplexing Metasurface

SEUNG-WOO NAM* and YOUNGJIN KIM*, Seoul National University, Republic of Korea
DONGYEON KIM, Seoul National University, Republic of Korea
YOONCHAN JEONG, Seoul National University, Republic of Korea

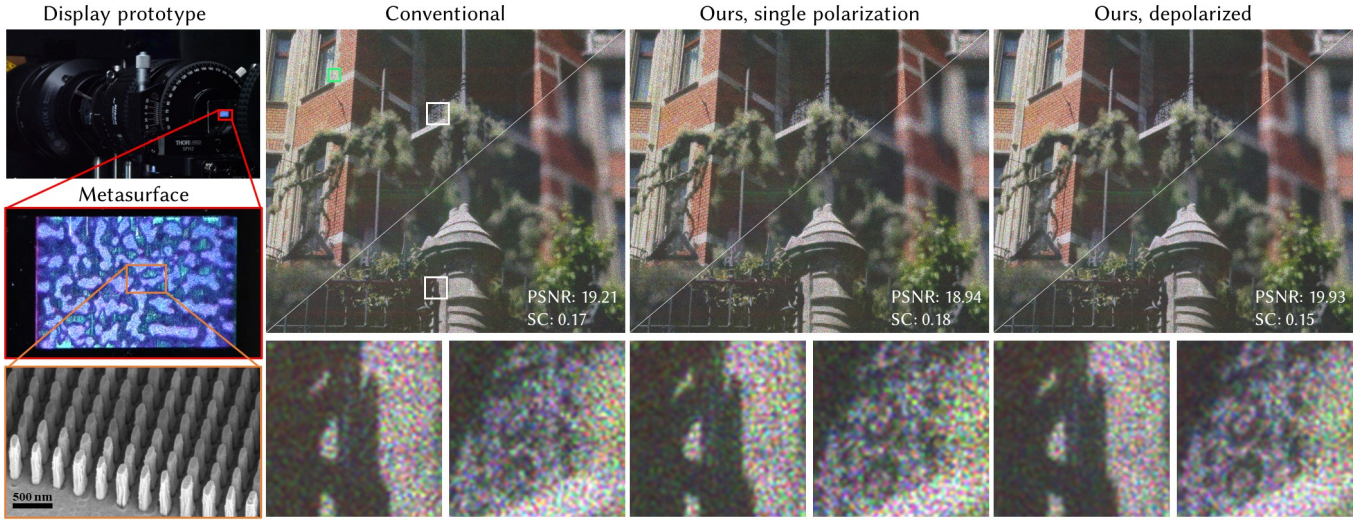

Fig. 1. (leftmost) Photographs of the depolarized holographic display prototype integrated with polarization-multiplexing metasurface. Gradual enlargements of the prototype show the metasurface and its scanning electron microscope (SEM) image. (right) Experimentally captured results with the display prototype. The first column presents the captured images from a conventional holographic display without a metasurface. The second column utilizes only the horizontal linear polarization channel of the metasurface. The last column shows the depolarized holography, in which both orthogonal polarization states are fully utilized. Mutual incoherence of orthogonal polarization states brought by the engineered metasurface enables the intensity sum without the interference of the reconstructed images from each polarization state. The speckle reduction effect arising from the incoherent nature of depolarization and the additionally offered degree of freedom by the jointly optimized metasurface significantly improve the image quality. The enlarged insets of our depolarized holography exhibit smoother speckle patterns, allowing high-frequency patterns in the image more visible compared to other cases. The green boxes specify area at which the speckle contrast is calculated. Source image credits to Kim et al. [2013].

The evolution of computer-generated holography (CGH) algorithms has prompted significant improvements in the performances of holographic displays. Nonetheless, they start to encounter a limited degree of freedom in CGH optimization and physical constraints stemming from the coherent nature of holograms. To surpass the physical limitations, we consider polarization as a new degree of freedom by utilizing a novel optical platform called metasurface. Polarization-multiplexing metasurfaces enable incoherent-like behavior in holographic displays due to the mutual incoherence of orthogonal polarization states. We leverage this unique characteristic of a metasurface by integrating it into a holographic display and exploiting polarization diversity to bring an additional degree of freedom for CGH algorithms. To minimize the speckle noise while maximizing the image quality, we devise a fully differentiable optimization pipeline by taking into account the metasurface proxy model, thereby jointly optimizing spatial light modulator phase patterns and geometric parameters of metasurface nanostructures.

*Both authors contributed equally to this research.

Authors' addresses: Seung-Woo Nam, 711asd@snu.ac.kr; Youngjin Kim, ttw8592@snu.ac.kr, Seoul National University, Republic of Korea; Dongyeon Kim, dongyeon93@snu.ac.kr; Yoonchan Jeong, Seoul National University, Republic of Korea, yoonchan@snu.ac.kr.

We evaluate the metasurface-enabled depolarized holography through simulations and experiments, demonstrating its ability to reduce speckle noise and enhance image quality.

## 1 INTRODUCTION

Recent breakthroughs in holographic displays have been primarily attributed to sophisticated computer-generated holography (CGH) algorithms, which have achieved exceptional image quality [Peng et al. 2020; Shi et al. 2021]. However, their technical advances have been gradually saturated as the performance of CGH algorithms is limited by the physical constraints of the display system. At this point, we raise a fundamental question: *Is there any degree of freedom unused in holographic displays?* Once the unexplored degree of freedom is identified, we can open up new possibilities for CGH algorithms to improve the performance of holographic displays.

The physical limitations present in holographic displays primarily stem from the coherent nature of light. Specifically, the speckle noise of the coherent light poses a challenge in holographic displays. The speckle noise not only degrades the image resolution [Deng and Chu 2017] but also hinders the accommodation response of the human eye [Kim et al. 2022b] induced by holographic stimuli. The utilization of a single static optical element for reducing speckle noise is highly advantageous, considering that conventional speckle reduction methods often involve sacrificing image resolution [Peng et al. 2021] or relying on time-averaged speckle intensity obtained from multiple frames using high-speed spatial light modulators (SLMs) [Lee et al. 2022]. In pursuit of this goal, we explore polarization as a novel optical channel that introduces incoherence and offers an additional degree of freedom to holographic displays.

Among the unique characteristics of light, polarization has been widely utilized in various imaging and display applications using two orthogonal polarization channels [Baek and Heide 2021; Hwang et al. 2022]. Particularly, the mutual incoherence of orthogonal polarization states is beneficial in holographic displays. However, it has been largely overlooked due to the lack of an appropriate optical platform for polarization-dependent modulation of light. Fortunately, a recently introduced optical element called metasurface, which modulates the optical response of light at the subwavelength regime [Khorasaninejad et al. 2016; Lin et al. 2014; Yu et al. 2011], provides a solution for this problem. Metasurfaces can offer uncorrelated phase profiles along the two orthogonal polarization states of the incident light, which is a unique characteristic hardly achieved with conventional optical devices [Arbabi et al. 2015a; Mueller et al. 2017]. Moreover, per-pixel modulation of optical response enables optimization-based design and makes them an ideal optical platform for holographic displays.

In this work, we propose a novel concept of holographic displays enabled by a polarization-multiplexing metasurface jointly designed with spatial light modulator phase patterns. Specifically, we employ a metasurface to exploit the polarization channel of the holographic display, generating two holograms with orthogonal polarization states simultaneously. We build a fully differentiable optimization pipeline to maximize the polarization-multiplexing functionality while considering the physical constraint of the metasurface. To this end, we model the electromagnetic response of the metasurface nanostructures to a differentiable proxy function and integrate it into a CGH optimization algorithm. The performance of our method is evaluated through simulations and experiments, demonstrating its competence in the overall image quality and speckle reduction compared to the conventional holographic displays. The quality improvement of holographic display, achieved through the joint engineering of static optical elements and SLM phase patterns, will undoubtedly open up a vast and exciting research field for the display community.

In summary, the major contributions of our work are as follows:

- We present a novel concept of holographic display which exploits orthogonal linear polarization states simultaneously. By incorporating a polarization-multiplexing metasurface, our approach expands the degree of freedom in holographic displays, making a room for CGH optimization algorithms.
- We devise an optimization pipeline that jointly optimizes the polarization-multiplexing metasurface and the SLM phase patterns. To the best of our knowledge, this is the first approach that uses a joint optimization method for co-designing a metasurface and a holographic display.
- We fabricate the optimized metasurface with electron beam lithography, and validate the proposed method through a benchtop prototype, verifying that experimental results are consistent with the simulations.

## 2 RELATED WORK

*Holography.* Holographic displays utilize interference of coherent light to reconstruct the object wavefront [Goodman 2005]. For their advantages in providing continuous depth cues, high-resolution images, and the aberration correction [Chang et al. 2020; Kim et al. 2021b; Nam et al. 2022; Park 2017], they have been adopted to near-eye displays in combination with various optical elements such as holographic optical elements [Li et al. 2016; Maimone et al. 2017; Yeom et al. 2015], geometric phase lenses [Kim et al. 2022a; Nam et al. 2020; Rous 2008], and waveguides [Jang et al. 2022]. In parallel with developments in display systems, CGH algorithms to design SLM phase patterns for desired images have also been developed. Numerous CGH algorithms that support various data types [Blinder et al. 2021; Chakravarthula et al. 2022b; Padmanaban et al. 2019; Shi et al. 2017] and optimization method for high-quality images [Chakravarthula et al. 2019; Fienup 1982; Gerchberg 1972; Zhang et al. 2017] have been proposed. Notably, computational methods that optimize parameterized real-world propagation models have achieved state-of-the-art results in experiments [Chakravarthula et al. 2020; Choi et al. 2021; Peng et al. 2020].

One of the major challenges to achieve high-quality images in holographic displays is the presence of speckle noise. The random phase distribution of the hologram introduces noise into the image, resulting in speckle intensity patterns. These speckle patterns, appearing as grainy textures, significantly reduce the quality of the image [Goodman 2007], particularly in the mid-high frequency range. The deterioration of the specific frequency region hinders the accommodation response induced by holographic stimuli, further limiting the realization of truly immersive 3D images [Kim et al. 2022b]. Efforts have been made to achieve speckle-free holographic displays through various approaches. Some methods involve using a partially-coherent light source [Deng and Chu 2017; Kozacki and Chlipala 2016; Lee et al. 2020; Peng et al. 2021], while others utilize high-speed SLMs to time-average multiple independent speckle patterns [Choi et al. 2022; Lee et al. 2022]. However, these approaches often face trade-offs between factors such as resolution, speckle contrast, depth of field and the number of time-multiplexed frames.

It is worth noting a recent work that has overcome the physical limitation of holographic displays, called étendue, by incorporating a random binary mask as an étendue expander [Kuo et al. 2020]. Though only validated in simulation, Baek et al. [2021b] extends this work and presents joint optimization of SLM phase patterns and complex-valued étendue expander with a large dataset. These works demonstrate the potential of breaking the physical constraints of holographic displays through optimized optical elements. While previous works have focused on utilizing small pixel pitches of additional optical elements for étendue expansion, our method takes advantage of the polarization-multiplexing characteristic of the metasurface to expand the degree of freedom in the polarization channel.

*Metasurface.* Metasurfaces are two-dimensional arrays of artificially designed nanostructures that modulate the diffraction of light at subwavelength regimes [Decker et al. 2015; Lin et al. 2014; Yu et al. 2011]. They attract much attention as the next generation of optical devices because they can realize optical functions that conventional refractive and diffractive optical devices cannot, such as wavelength-multiplexing [Arbabi et al. 2018; Li et al. 2021; Shi et al. 2018], angle-multiplexing [Jang et al. 2021; Kamali et al. 2017], and complex (amplitude and phase) modulation [Lee et al. 2017; Overvig et al. 2019]. The lithography-based nanofabrication process shows the potential for mass production utilizing the current foundry legacy. Utilizing the advantages of these metasurfaces, many applications have been reported such as flat lenses [Arbabi et al. 2015b; Khorasaninejad et al. 2016; Kim et al. 2020, 2021a], beam shaping [Sroor et al. 2020], holography [Huang et al. 2013; Zheng et al. 2015], optical filters [Zhou et al. 2020; Zou et al. 2022], and biosensing [Leitis et al. 2019; Yavas et al. 2017; Yesilkoy et al. 2019]. Additional information about the basic principles, fabrication, and applications are located in the review articles [Chen et al. 2016; Hsiao et al. 2017; Lee et al. 2019].

Among the benefits of metasurfaces, the ability to control electromagnetic waves independently for two orthogonal polarization states is a powerful feature unique to metasurfaces. Arbabi et al. [2015a] and Mueller et al. [2017] theoretically and experimentally demonstrated that by adjusting the geometric dimensions and rotation angle of the nanostructures comprising the metasurface, it is possible to independently control the phases of light for two arbitrary orthogonal elliptical polarization states [Guo et al. 2019]. Rubin et al. [2019; 2021] generalizes the metasurface design strategy using Jones matrix calculation.

Unlike conventional diffractive optical elements, metasurfaces should be designed considering the electromagnetic response at the nanoscale, as the nanostructures are arranged with a period smaller than the wavelength of the incident wave. The most common process in designing a metasurface utilizes the pre-simulated optical response library of nanostructures that can be used as a look-up table to specify the geometric dimensions [Chen et al. 2016]. However, this method does not reflect the electromagnetic effect during the optimization step and thereby cannot consider the interactions among adjacent nanostructures or the limited degree of freedom under physical constraints. The ideal solution to handle this problem is a full-field simulation [Chung and Miller 2020; Mansouree et al. 2020; Steinberg and Yan 2021], but these are computationally expensive, being impractical for designing the aperiodic metasurface even with several hundreds of micrometers. In contrast to that, some works introduce the proxy model that approximates the solutions of the Maxwell equation, thereby building the differentiable pipeline under consideration of nanoscale optical response [Pestourie et al. 2018]. Research on large-scale metalens designed with a metasurface proxy model shows the possibility of practical use in the virtual-reality system [Li et al. 2022].

*Computational optics design.* In the field of computational photography, the joint optimization of the optical component and the post-processing algorithm in an end-to-end manner has been extensively explored and has demonstrated its effectiveness in domain-specific imaging systems. These end-to-end frameworks enable the differentiable optimization of optical components, such as binary masks [Chakrabarti 2016; Iliadis et al. 2020], diamond-turned refractive optics [Peng et al. 2019], compound optics [Tseng et al. 2021b], and diffractive optical elements [Shi et al. 2022], in conjunction with post-processing algorithms using large datasets. They have achieved state-of-the-art results in various applications, including extend depth of field and super-resolution imaging [Sitzmann et al. 2018], super-resolution SPAD imaging [Sun et al. 2020b], hyperspectral imaging [Baek et al. 2021a; Dun et al. 2020], depth sensing [Chang and Wetzstein 2019; Haim et al. 2018; Wu et al. 2019], and high dynamic range imaging [Metzler et al. 2020; Sun et al. 2020a].

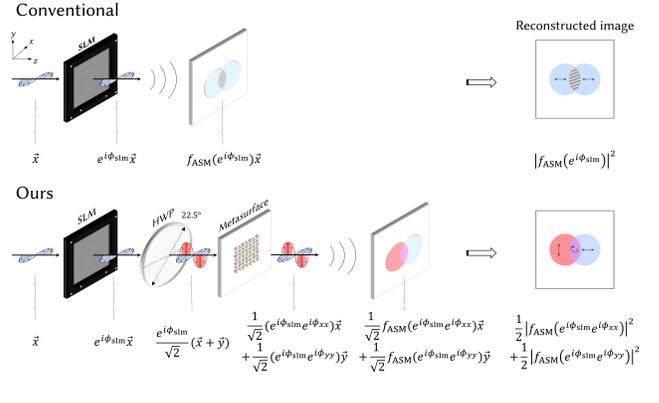

Fig. 2. Conceptual schematics and the field evolution along the optical system of the conventional and the proposed holographic display. In a traditional system, (most of) the phase-only SLM imparts the phase profile ($\phi_{\text{slm}}$) for the linearly polarized incident light. Coherence of light source results in interference among wavefronts. In contrast to that, our scheme utilizes the orthogonal polarization states using metasurface. After passing through the SLM, the HWP rotates the horizontally polarized light by 45 degrees, making diagonal linear polarization state. Then, the polarization-multiplexing metasurface provides different phase modulation for each linear polarization state. Consequent reconstructed fields along two orthogonal polarization channels do not interfere with each other, thereby leading to a weighted intensity sum.

Most recently, researchers have applied metasurfaces to the end-to-end optimization framework described above to achieve unprecedented functionalities. Tseng et al. [2021a] proposed a joint optimization of single metalens and the decoding neural network that achieves a high-quality imaging performance within an extremely small form factor. Hazineh et al. [2022] suggested single-shot depth sensing and spatial frequency filtering metalenses based on a TensorFlow framework. Trained metasurface resonator encoders for real-time hyperspectral imaging have also been reported [Makarenko et al. 2022]. Similar to aforementioned works, we extend the joint optimization approach to holographic displays, aiming to expand the degree of freedom in CGH optimization and improve the image quality. We present a differentiable optimization algorithm that can jointly optimize the metasurface geometric parameters and the SLM phase pattern.

## 3 METHODS

### 3.1 Preliminaries

We first briefly define the notation used to describe the wave propagation model that includes polarization. Throughout this paper, we describe the polarization of the light using the Jones calculus, where the polarization state are denoted with 2×1 Jones vectors and optical elements are described with 2×2 Jones matrices.

$$\vec{v} = \begin{bmatrix} v_x \\ v_y \end{bmatrix}, \quad J = \begin{bmatrix} J_{xx} & J_{xy} \\ J_{yx} & J_{yy} \end{bmatrix}. \quad (1)$$

Based on this notation, we define horizontal linear polarization as $\vec{x} = [1, 0]^\top$ and vertical linear polarization as $\vec{y} = [0, 1]^\top$. Therefore, elements $v_x, v_y$ of Jones vector $\vec{v}$ are complex-valued amplitude of horizontal and vertical linear polarization components. The polarization operation of an optical element is calculated by matrix multiplication between the Jones matrix and the Jones vector. We use (·) to represent matrix multiplication.

### 3.2 Depolarized holography

We exploit polarization in holographic displays by depolarizing the light using the metasurface. Depolarization itself is not a novel idea and has already been used in digital holography and holographic projection for speckle suppression [Bianco et al. 2018; Goodman 2007; Rong et al. 2010]. However, this method relies on the random behavior of a diffusive screen that is difficult to be delicately designed. Instead, we use a polarization-multiplexed metasurface that allows per-pixel control of phase modulation. This design flexibility of the metasurface offers potential for further improvements in performance. Additionally, the metasurface with known amplitude and phase enables CGH optimization incorporating the metasurface.

Figure 2 illustrates the comparison between conventional holographic displays and the proposed method. Conceptually, a holographic display can be abstracted to a simple optical system consisting of a laser and an SLM. Free-space wave propagation in holography is described in the angular spectrum method (ASM) [Matsushima and Shimobaba 2009] expressed as

$$f_{\text{ASM}}(u, z) = \mathcal{F}^{-1}\left\{\mathcal{F}\{u\} \mathcal{H}(v_x, v_y, \lambda, z)\right\}$$

$$\mathcal{H}(v_x, v_y, \lambda, z) = \begin{cases} e^{i2\pi z \sqrt{1/\lambda^2 - v_x^2 - v_y^2}}, & \text{if } \sqrt{v_x^2 + v_y^2} < \frac{1}{\lambda} \\ 0, & \text{otherwise} \end{cases} \quad (2)$$

where $u$ is a complex-valued wavefront, $z$ is a propagation distance, $\lambda$ is a wavelength, and $v_x, v_y$ are spatial frequencies. For the sake of simplicity, we omit $\lambda$ for the rest of the equations. Generally, CGH optimization algorithms calculate the propagated field through the ASM and iteratively optimize the amplitude of the propagated field to be the target amplitude.

In our method, a half-wave plate (HWP) and a metasurface are placed after the SLM. The angle between a horizontal line and the fast axis of the HWP is set to 22.5 degrees, so our HWP rotates the incident horizontally polarized light from SLM by 45 degrees, making a diagonal polarization state. Since the diagonal polarization state can be separated into a horizontal and vertical linear polarization with identical amplitude, the metasurface after the HWP applies different phase modulation on these orthogonal linear polarization states. The Jones matrix of the HWP and the metasurface are described as

$$J_{\text{hwp}} = \frac{1}{\sqrt{2}} \begin{bmatrix} 1 & -1 \\ 1 & 1 \end{bmatrix}, \quad J_{\text{meta}} = \begin{bmatrix} e^{i\phi_{xx}} & 0 \\ 0 & e^{i\phi_{yy}} \end{bmatrix}, \quad (3)$$

where $\phi_{xx}, \phi_{yy}$ are the phase shifts on the co-polarized component of transmitted light. We assume that the SLM, the HWP, and the metasurface are sufficiently close to be located on the same plane. The Jones vector of the complex-valued wavefront after the metasurface is expressed as a matrix multiplication of the Jones vector of the SLM field ($e^{i\phi_{\text{slm}}}\vec{x}$), the Jones matrix of HWP ($J_{\text{hwp}}$), and that of metasurface ($J_{\text{meta}}$). Therefore, the complex-valued wavefront at distance $z$ of the proposed system and the intensity of the corresponding field are expressed as

$$f_{\text{depol}}(\phi_{\text{slm}}, z) = f_{\text{ASM}}\left(J_{\text{meta}} \cdot J_{\text{hwp}} \cdot e^{i\phi_{\text{slm}}}\vec{x}, z\right),$$

$$\left|f_{\text{depol}}(\phi_{\text{slm}}, z)\right|^2 = \frac{1}{2} \sum_{p \in (x,y)} \left|f_{\text{ASM}}\left(e^{i\phi_{pp}} e^{i\phi_{\text{slm}}}, z\right)\right|^2. \quad (4)$$

Here, the intensity of the propagated field is expressed as the intensity sum of the field evolved with two orthogonal polarization states due to mutual incoherence. Therefore, the intensity of the propagated field resembles that of the partially-coherent holographic displays, where polarization diversity replaces previously exploited diversities: angle diversity [Lee et al. 2020], wavelength diversity [Deng and Chu 2017; Kozacki and Chlipala 2016; Peng et al. 2021], and time-multiplexed frames [Choi et al. 2022; Lee et al. 2022].

When the hologram generated from the SLM phase pattern is depolarized into two linear orthogonal polarization channels as it passes through the metasurface, the holographic images of each polarization should be complementary to each other to improve the image quality. Therefore, the optimization of the phase distribution $\phi_{xx}, \phi_{yy}$ of the metasurface for the two orthogonal polarization states is the core of this work, which determines the performance of the depolarized holography. This necessitates a deliberate metasurface design through optimization.

### 3.3 Joint optimization pipeline for polarization-multiplexing metasurface design

*Metasurface proxy model.* We use a linear polarization basis for the metasurface design since it rarely introduces an undesirable cross-polarization leakage in the multi-wavelength regime [Arbabi et al. 2015a; Mueller et al. 2017]. However, it is difficult for the silicon nitride metasurfaces to fully cover the $2\pi$ radian range of phase modulation on both orthogonal polarization states under the practical fabrication conditions, due to the low refractive index (see section S1.1 in Supplementary Material for more details). Additionally, the dispersion characteristic of the dielectric material results in varying phase shifts for different wavelengths. To take account of these issues, we adopt the differentiable metasurface proxy model to solve the physically constrained problem stemming from the phase modulation range and material dispersion.

The establishment of the proxy model is divided into three major steps. First, the electromagnetic response of the nanostructures is simulated by rigorous coupled-wave analysis (RCWA) [Kim and

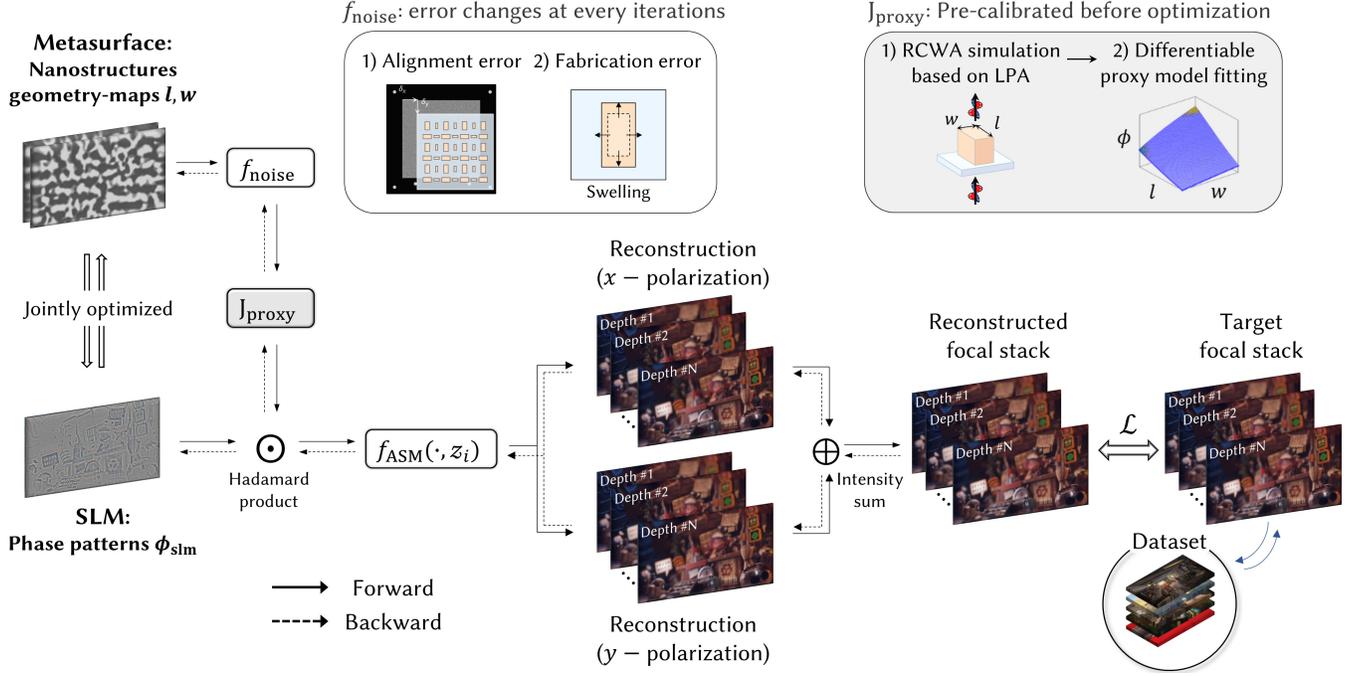

Fig. 3. Illustration of the joint optimization pipeline. Nanostructure geometry-maps of the metasurface are jointly optimized with the SLM phase pattern to realize the focal stack holograms over the target image dataset. The SLM field evolves into two distinct holograms by the polarization multiplexing metasurface. In this process, the noise-reflected Jones matrix models the optical operation of the metasurface under experimental situations. Two holograms for each polarization state propagate to all target planes, then combined by intensity summation for each depth. Backpropagating gradients of the loss calculated between the reconstructed and target focal stacks updates the metasurface and SLM phase patterns. Source image credits to Alex Treviño.

Lee 2023] under local periodic approximation (LPA) [Li et al. 2022; Pestourie et al. 2018; Tseng et al. 2021a]. Given that the pixel pitch of the metasurface and the height of the nanorod are decided, we obtain the modulated phase as a function of the length and width $(l, w)$ of the nanorod. The combination of two polarization states $(\phi_{xx}, \phi_{yy})$ and wavelengths (638, 520, and 450 nm) results in a total of six libraries. Next, for each wavelength, the libraries of $\phi_{xx}, \phi_{yy}$ are fitted by linear quadratic polynomials and used to represent the Jones matrix, of which the general formulation can be written as follows:

$$\mathrm{J}_{\mathrm{proxy}}(l, w) = \begin{bmatrix} e^{i \sum_{n,m=0}^{2} c_{nm} l^n w^m} & 0 \\ 0 & e^{i \sum_{n,m=0}^{2} \tilde{c}_{nm} l^n w^m} \end{bmatrix} \quad (5)$$

where $l, w$ are normalized by the pixel pitch of the metasurface, $c_{nm}$, $\tilde{c}_{nm}$ are the polynomial coefficients, and $\mathrm{J}_{\mathrm{proxy}}$ is an approximated Jones matrix of the metasurface. More details about the libraries and fitted polynomials can be found in the Supplementary Material.

*Joint optimization pipeline.* While the metasurface can be engineered for our depolarized holography, the SLM phase patterns can also be optimized for the metasurface. Therefore, we jointly optimize the metasurface and SLM phase patterns. Figure 3 illustrates our joint optimization pipeline. The proposed pipeline is based on the CGH optimization algorithm with focal stack supervision [Choi et al. 2022; Lee et al. 2022]. We choose a focal stack as an optimization target since it is a challenging, over-constrained problem for a single SLM phase pattern in conventional holographic displays. We evaluate the degree of freedom brought by the metasurface and the joint optimization through focal stack holograms.

In our pipeline, we jointly optimize two parameters: the geometry-maps of the metasurface and SLM phase patterns. First, the complex-valued amplitude of the metasurface is calculated from the geometry-maps using the pre-calibrated metasurface proxy model in Equation 5. In addition, we implement a noise function $f_{\mathrm{noise}}$ to simulate the potential alignment and fabrication errors that possibly occur during real-world experiments, thereby making the optimized metasurface robust against these imperfections.

$$f_{\mathrm{noise}}(l(x)) = l(x) * \delta(x - x_\epsilon) + l_\epsilon. \quad (6)$$

In the equation, $*$ is convolution, and $\delta(\cdot)$ represents the Dirac delta function. The metasurface is shifted by misalignment noise $x_\epsilon$ determined by uniform random distribution $\mathcal{U}(-\sigma_x, \sigma_x)$, and absolute value of Gaussian noise $l_\epsilon \sim |\sigma_l^2 \mathcal{N}(0, 1)|$ is added for the fabrication error. Though we only express dependencies in $l$ for dimension $x$ in the equation, the same applies to parameter $w$ and dimension $y$.

With the noise function in Equation 6, we can express the noise-reflected Jones matrix of the metasurface $\mathrm{J}_{\mathrm{proxy}}$ by substituting it into the Equation 5. Therefore, the amplitude of the propagated field is obtained with Equation 4, and we compare it with the target

amplitude as

$$\min_{\{l,w,\phi\}} \mathcal{L}\left(\left|f_{\text{ASM}}\left(J_{\text{proxy}}\left(f_{\text{noise}}(l,w)\right) \cdot J_{\text{hwp}} \cdot e^{i\phi}\vec{x}, z^{\{d\}}\right)\right|, a_{\text{target}}^{\{d\}}\right), \quad (7)$$

where $\mathcal{L}$ is a loss function, $\phi$ is an SLM phase pattern, and $\{d\}$, $d = 1 \ldots D$ is the index of the propagation distances. We optimize the geometry-maps $(l, w)$ of the metasurface through a large dataset where SLM phase patterns $\phi$ are optimized per each target image.

Algorithm 1 demonstrates the algorithm used for the metasurface optimization in detail. We alternately update the metasurface geometry-maps and the SLM phase patterns using the stochastic gradient descent method. During the metasurface optimization, alignment and fabrication errors are simulated with the noise function $f_{\text{noise}}$ and applied to the metasurface, and the geometry-maps are updated to minimize the loss defined in the Equation 7. However, the SLM phase patterns cannot converge to a certain solution if the position and values of the metasurface geometry-maps change every iteration. Therefore, we leave out the noise function $f_{\text{noise}}$ during the phase pattern optimization and assume the ideal metasurface profile without fabrication and alignment errors. To the best of our knowledge, a high-resolution RGB-D dataset of natural images does not exist, so we use the DIV2K dataset [Agustsson and Timofte 2017] and generate focal stacks from a single 2D image as target data. For every training data, a 2D image is placed at a randomly selected plane, and the focal stack of incoherent propagation is calculated from the image [Lee et al. 2022].

We implement our algorithm in PyTorch and utilize the automatic differentiation tools for joint optimization. The metasurface is trained for 2000 epochs with 100 data samples in the DIV2K train set. We set the learning rate for the SLM phase patterns to $1e^{-1}$, and for the metasurface to $5e^{-3}$. The SLM phase patterns are initialized as a uniform random phase from the range of $[-\pi, \pi]$, while the metasurface geometry-maps are initialized with a uniform random distribution from the range of $[-1e^{-3}, 1e^{-3}]$. The joint optimization takes approximately 37 hours to converge on an NVIDIA RTX A6000. Our source code is available on the project website.

## 4 SIMULATION RESULTS

Throughout the paper, as the target for the metasurface optimization is a focal stack, we evaluate our method with focal stack holograms generated from either 2D images or RGB-D data. To evaluate the image quality, we primarily utilize two metrics: peak signal-to-noise ratio (PSNR) and speckle contrast (SC). PSNR is calculated as the mean squared error between reconstructed images and target images across all 7 depth planes, encompassing both the focused and defocused images. This provides an assessment of the overall image quality of the focal stack. SC serves to quantify the presence of speckle noise in the image, representing the extent of intensity fluctuations of a speckle pattern relative to the average intensity. SC is defined by

$$SC = \frac{\sigma_I}{\bar{I}}, \quad (8)$$

where $\sigma_I$ and $\bar{I}$ represent the standard deviation and average of the intensity, respectively. SC ranges from 0 to 1, where a value of 1 indicates fully developed speckles, while lower SC values correspond to reduced speckle noise. In practice, the maximum value of SC is

---

**Algorithm 1:** Joint optimization pipeline

$E$ : Number of epochs
$N$ : Number of training data
$\mathcal{L}$ : Loss function
$\alpha_{\text{meta}}, \alpha_{\text{slm}}$ : Learning rates

**for** $e$ in $1 \ldots E$ **do**
    **for** $n$ in $1 \ldots N$ **do**
        // Metasurface optimization
        $J_{\text{meta}} \leftarrow J_{\text{proxy}}(f_{\text{noise}}(l, w))$
        $u_{\text{meta},n} \leftarrow J_{\text{meta}} \cdot J_{\text{hwp}} \cdot e^{i\phi_{n,e}}\vec{x}$
        $a_{\text{recon},n}^{\{d\}} \leftarrow \left|f_{\text{ASM}}\left(u_{\text{meta},n}, z^{\{d\}}\right)\right|$
        $(l, w) \leftarrow (l, w) - \alpha_{\text{meta}} \cdot \mathcal{L}\left(a_{\text{recon},n}^{\{d\}}, a_{\text{target},n}^{\{d\}}\right)$

        // SLM phase pattern optimization
        $J_{\text{meta}} \leftarrow J_{\text{proxy}}(l, w)$
        $u_{\text{meta},n} \leftarrow J_{\text{meta}} \cdot J_{\text{hwp}} \cdot e^{i\phi_{n,e}}\vec{x}$
        $a_{\text{recon},n}^{\{d\}} \leftarrow \left|f_{\text{ASM}}\left(u_{\text{meta},n}, z^{\{d\}}\right)\right|$
        $\phi_{n,e} \leftarrow \phi_{n,e} - \alpha_{\text{slm}} \cdot \mathcal{L}\left(a_{\text{recon},n}^{\{d\}}, a_{\text{target},n}^{\{d\}}\right)$
        $\rightarrow$ save updated $\phi_{n,e}$
    **end**
**end**
**return** $(l, w)$

---

not 1 even with the fully developed speckles due to the influence of the optical system and the image sensor, which determine the number of independent phasor arrays [Goodman 2007]. We provide the speckle contrast estimated in simulation and measured with the experimental setup using identical settings. The speckle contrast is measured in the selected area whose intensity distribution is uniform, which is indicated by the green box in the figures. By utilizing these two image quality metrics, we analyze the advantages of the polarization-multiplexing metasurface in two aspects: providing an additional degree of freedom for focal stack hologram optimization and speckle reduction.

*Tolerance in imperfections.* Figure 4 illustrates the impact of employing the noise function during the metasurface optimization. We optimize the metasurface with and without the noise function and compare the effect of misalignment and fabrication in these two metasurfaces. The upper left sub-images represent the reconstructed images assuming flawless fabrication and perfect alignment of the system. The lower-right sub-images show the reconstructed image under mismatched conditions; the metasurface is shifted 10 pixels (31 $\mu$m) horizontally and vertically from the SLM, and a fabrication error is introduced in the geometry-maps, following a Gaussian distribution with a deviation of 5 nm. It is evident that the existence of noise function in the entire optimization pipeline effectively mitigates the impact of imperfections, making the designed metasurface more practical for real-world applications. Notably, even the PSNR of the case without mismatch increases with the utilization of the noise function. We presume that the noise function also prevents the overfitting of the metasurface profile to the training dataset.

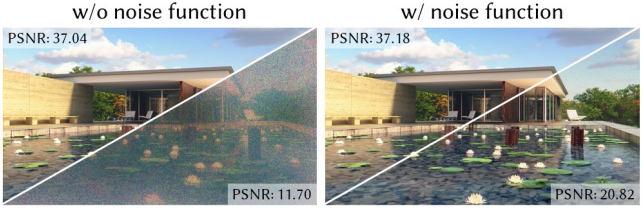

Fig. 4. Simulation results for validating the effect of the noise function. The upper left sub-images depict the image obtained when the metasurface is precisely aligned and fabricated without error, resulting in the reconstruction of phase patterns in the identical setting as CGH optimization. In contrast, lower right sub-images show the reconstructed results when the metasurface is misaligned by 10 pixels horizontally and vertically, accompanied by a fabrication error. Even with these slight errors, the image quality is severely degraded in the absence of the noise function during the metasurface optimization. Source image credits to eMirage.

*Focal stack holograms.* We compare the simulation results of a total of four scenarios for the evaluation: a conventional holographic display without a metasurface (*conventional*), a depolarized holography with a metasurface fabricated from a geometry-map of a random distribution (*random depol*), and with the optimized metasurface utilizing only a single polarization state (*optimized single-pol*) or depolarized with a diagonal polarization (*optimized depol*). We include the *random depol* case in our simulation as a baseline of polarization-multiplexing metasurface without optimization. By comparing the random metasurface and the optimized metasurface, we can distinguish the effect of the depolarized holography and the joint optimization pipeline.

The simulated results in Figure 5(a) show that both holograms optimized to focal stacks generated from 2D (first row) and RGB-D (second row) data exhibit a reduction in speckle noise when the metasurface is inserted and depolarized, regardless of whether it is optimized or not. When comparing the two depolarized metasurfaces, *optimized depol* outperforms *random depol* in terms of PSNR and speckle contrast, demonstrating the effectiveness of metasurface optimization. The case of *optimized single-pol* provides insight about how our polarization-multiplexing metasurface works. Even with the optimized metasurface, the image quality is worse than *conventional*, if only a single polarization state is available. The polarization-multiplexing metasurface reconstructs two slightly different 'worse holograms', and the incoherent summation of these two holograms by depolarizing results in the best image quality, which is equivalent to *optimized depol*. It is worth noting that even though the metasurface is optimized with focal stacks generated from 2D images, focal stacks from RGB-D also shows improvements. This suggests that our joint optimization pipeline is generalized enough to be used in other types of holograms that are not specifically used during the optimization process.

Figure 5(b) presents the quantitative analysis of the four scenarios. We draw a histogram from the region specified by the green box in the first row of Figure 5(a). The black dashed line on the histogram indicates the peak of the intensity distribution of the ground truth image. Since the selected region has nearly uniform intensity, a sharp peak centered around the black dashed line implies the intensity distribution close to the ground truth. It is clear that *optimized depol* has the sharpest intensity distribution among all cases, indicating reduced speckle noise without compromising image contrast. While the peak of *conventional* and *optimized depol* are close to the ground truth, the histogram shows broader distributions due to severe speckle noise. However, *random depol* exhibits a slight shift in the peak, failing to accurately reproduce the intensity of the ground truth image. This results in a low-contrast image, as observed in Figure 5(a). Additionally, we include a graph of the average PSNR and speckle contrast of focal stack holograms obtained from 30 natural images in the DIV2K validation set [Agustsson and Timofte 2017], which are not used during the metasurface optimization. The graph confirms that our previous observations in Figure 5(a) apply to general images; our depolarized holography outperforms the conventional method by 4.36 dB through the incoherent superposition of two noisy holograms.

*Understanding the optimized metasurface.* Though the optimization of the metasurface enhances the image contrast and reduces speckle noise, there is a trade-off due to the limited degree of freedom that a single metasurface can provide. Figure 6 demonstrates the simulation results highlighting the disadvantage of the optimized metasurface compared to the random metasurface when generating independent images for two orthogonal polarization states. In this simulation, we optimize a single SLM phase pattern to generate different images for vertical and horizontal polarization state, thereby indirectly assessing the ability of the metasurface to control these polarization states independently. The random metasurface outperforms the optimized metasurface in generating polarization-dependent images, contrary to the case of focal stack generation. Hence, we conclude that our joint optimization pipeline tailors the randomness of the metasurface to maintain image contrast while ensuring that the two holograms in orthogonal polarization states are sufficiently distinct to benefit from incoherent superposition.

## 5 EXPERIMENT

### 5.1 Implementation

*Metasurface fabrication.* The metasurface is fabricated utilizing electron beam lithography according to the flowchart sequence shown in Figure 7(a). A 0.5 mm thick glass wafer is cleaned with a sulfuric acid peroxide mixture (SPM), followed by 800 nm deposition of silicon nitride (SiN) utilizing a plasma-enhanced chemical vapor deposition equipment (P5000, AMAT). Two layers of electron beam resist are then spin-coated onto the SiN layer. First, a PMMA 495A4 solution is spin-coated at 500 rpm for 5 seconds and 2000 rpm for 40 seconds, followed by soft-baking at 180 degrees for 3 minutes. Then a PMMA 950A2 solution is spin-coated at 500 rpm for 5 seconds and 3000 rpm for 40 seconds. Similarly, the soft-bake process is done at 180 degrees for 3 minutes. To prevent charge accumulation issues during electron beam lithography, the conducting polymer (ESPACER 300Z, SHOWA DENKO) is spin-coated at 500 rpm for 5 seconds and 2000 rpm for 30 seconds. The designed nanopatterns are produced using electron beam lithography (JBX-6300FS, JEOL), which takes about 20 hours to fabricate two metasurfaces for experimental demonstration. After exposure, water-soluble conducting polymer is removed with DI water and the resist layers are developed by soaking the sample in the development solution (MIBK:IPA=1:3, MICROCHEM) for 3 minutes. Chromium (Cr) with a thickness of 40 nm is then deposited using an electron beam evaporator and use aceton to lift off the PMMA resist layers to complete the hardmask patterning. After the SiN etching process (ICP 380, OXFORD SYSTEM100), the remaining Cr hardmask is removed with Cr etchant (CE-905N, TRANSENE), and the desired metasurface is finally fabricated, as shown in Figure 7(b).

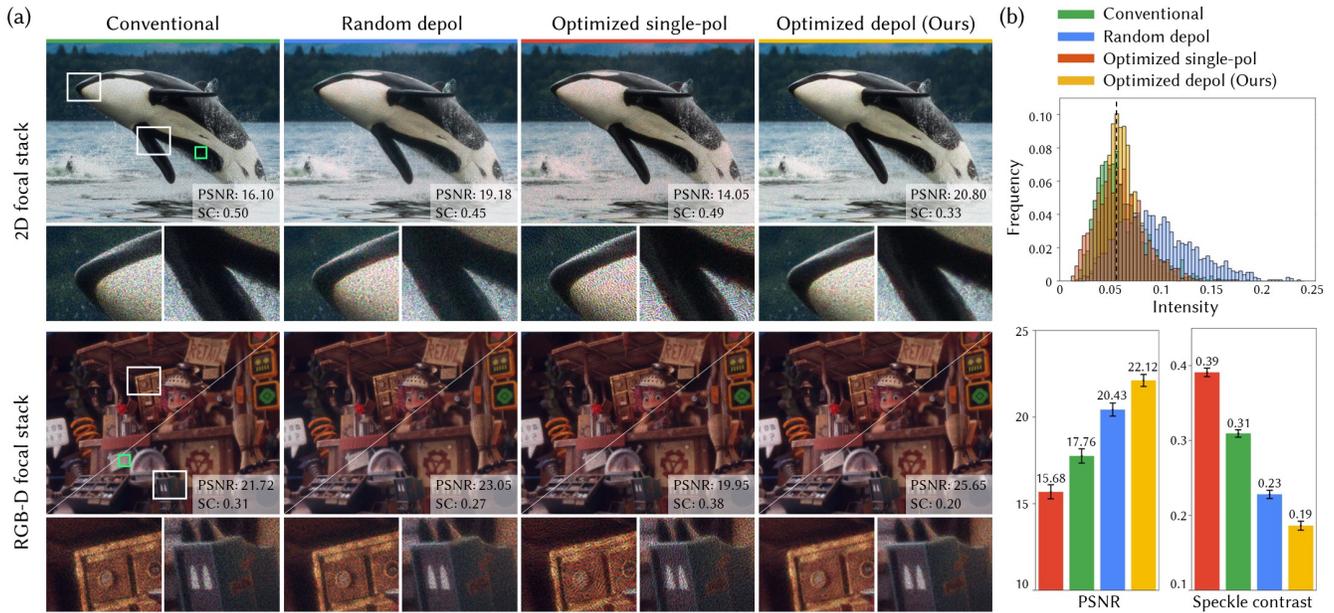

Fig. 5. Evaluation results in simulation. (a) The four columns correspond to the reconstructed images in the following cases: without a metasurface (*conventional*), depolarized with a metasurface of randomized geometry-maps (*random depol*), with an optimized metasurface utilizing a single polarization (*optimized single-pol*), and depolarized with an optimized metasurface (*optimized depol*), from left to right. The first and second rows display focal stack images derived from 2D and RGB-D data, respectively. In the second row, the upper left sub-images display the image focused at the far plane, while the lower right sub-images show the image focused at the near plane. The PSNR and speckle contrast are provided in the lower right of each image. The green boxes indicate the specified area where the speckle contrast is calculated. The orca image credits to Wirestock Creators and the Junk Shop image credits to Alex Treviño. (b) (upper) Image histogram of the region indicated by the green box in the figures of the first row in (a), with the most frequent intensity of the ground truth image marked by a dashed black line. The averaged PSNR (lower left) and speckle contrast (lower right) of the focal stack holograms are calculated from 30 natural images from DIV2K validation dataset [Agustsson and Timofte 2017], with error bars indicating a standard error.

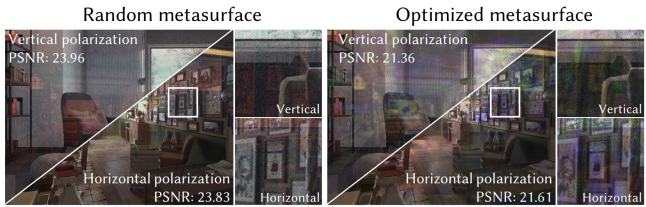

Fig. 6. Simulation results demonstrating polarization-dependent image generation. A single SLM phase pattern is optimized to generate distinct 2D images for each orthogonal polarization states with a metasurface specified above the figure. Compared to the random metasurface, the reconstructed images of each polarization state utilizing the optimized metasurface show significant noise in the image and are more challenging to differentiate. Source images credit to Flavio Della Tommasa and Blender Animation Studios.

*Display system.* We evaluate our method using a benchtop holographic display prototype. A collimated laser (FISBA READYBeam) incidents to the SLM (HOLOEYE LETO-3) and passes through the $4f$ system, which filters out the high order diffraction terms and relays the wavefront of the SLM. We note here that our $4f$ system demagnifies the SLM with a magnification factor of approximately 0.5 to match the size of the SLM and the metasurface, which is fabricated to dimensions of 3.4×6.0 mm$^2$. Following the $4f$ system, an HWP (Thorlabs AHWP10M-600) is placed to rotate the direction of linear polarization. The fabricated metasurface is carefully placed after the HWP, aligned with the relayed SLM. An additional $4f$ system is employed after the metasurface to image the SLM plane for the alignment of the metasurface and the SLM. Once alignment is achieved, this second $4f$ system is no longer required. Reconstructed images are captured from multiple planes using a CCD (FLIR GS3-U3-51S5M-C) mounted on a motorized stage (Newport FCL100). The schematic diagram of the benchtop prototype is illustrated in Figure 13(a).

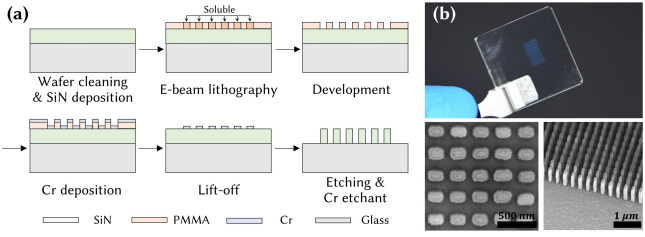

Fig. 7. (a) Metasurface fabrication flowchart. (b) Metasurface fabrication results. The fabricated metasurface measures approximately 3.4×6.0 mm². Photograph of the metasurface and its SEM images along the top-view and tilt-view, respectively.

*Metasurface alignment.* Precise alignment is crucial for our system since we optimize the metasurface under the assumption that the SLM and the metasurface are in the same position. Our alignment procedure can be divided into two main steps. First, we align the metasurface in 3 axes using two motorized stages and a manual stage. This is done by imaging the SLM plane through the second $4f$ system and observing the boundary lines of the SLM and the metasurface. Second, we perform camera-in-the-loop (CITL) model calibration and learn the actual phase and amplitude of the fabricated metasurface. This step is similar to a fine-tuning step of post-processing algorithms performed in many end-to-end cameras [Shi et al. 2022; Tseng et al. 2021a]. Through CITL model calibration, both misalignment and fabrication error of the metasurface are measured and included in the CGH optimization process.

We note here that our calibration process is relatively simpler compared to that of Kuo et al. [2020], which aligned the SLM and the random binary mask using optimized SLM phase patterns that generate a single focal spot. Since we optimize the metasurface with the noise function $f_{\text{noise}}$ for alignment robustness, the metasurface phase pattern is intentionally designed to be coarse enough to be less sensitive to alignment errors. The effectiveness of the noise function in achieving alignment robustness is already discussed in Section 4. Furthermore, these coarse phase patterns allow us to calibrate the metasurface phase and amplitude through CITL optimization. Additional information regarding the alignment of the metasurface and the SLM can be found in the Supplementary Material.

*Model calibration with camera-in-the-loop training.* We use a CITL-calibrated wave propagation model during the experimental validation [Choi et al. 2022, 2021; Jang et al. 2022; Peng et al. 2020]. The CITL-calibrated model helps to reduce the discrepancy between the simulation and the real-world, thereby providing a clearer evaluation of the proposed method. In order to accurately capture the physical phenomenon of polarization-multiplexing, we combine the deep neural network-based propagation model developed by Choi et al. [2022] and the all-physically interpretable model introduced by Jang et al. [2022], with slight modifications to incorporate the Jones matrices of the HWP and the metasurface.

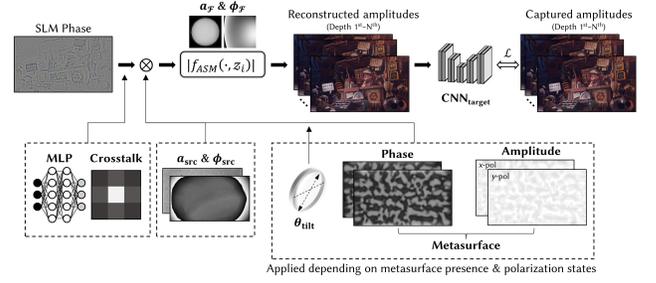

Fig. 8. Schematic illustration of our wave propagation model CITL calibration framework. Our model includes the light source, SLM, Fourier plane, HWP, and the metasurface, which are parameterized to account for the contents-independent terms. Additionally, a convolutional neural network is incorporated for the contents-dependent terms. The propagation model is trained with a dataset of captured amplitudes. Source image credits to Alex Treviño.

The schematic diagram of the proposed model is depicted in Figure 8. A multi layer perceptron (MLP) and a 3×3 kernel $k$ model spatially-varying phase response and crosstalk between adjacent SLM pixels. Source intensity $a_{\text{src}}$, phase $\phi_{\text{src}}$ and complex field of Fourier plane $a_{\mathcal{F}}, \phi_{\mathcal{F}}$ are incorporated into ASM to account for contents-independent propagation terms. Different from other models, we introduce the metasurface and the HWP to characterize our depolarized holography. The amplitude and phase of the two polarization states of the metasurface are learned to reflect the actual fabrication and alignment results. The rotation angle $\theta_{\text{tilt}}$ of the HWP is parameterized to consider the mismatch between the polarization direction of the light source and the fast axis of the HWP. The reconstructed amplitudes then pass through a convolutional neural network $\text{CNN}_{\text{target}}$, that incorporates contents-dependent terms. To conclude, our propagation model is expressed as

$$f_{\text{model}}(\phi) = \text{CNN}_{\text{target}}\left(f_{\text{ASM}}\left(J_{\text{proxy}}(l,w) \cdot J_{\text{hwp}}(\theta_{\text{tilt}})\right.\right. \\ \left.\left. \cdot a_{\text{src}} e^{i\phi_{\text{src}}} e^{i(k*\text{MLP}(\phi))}; a_{\mathcal{F}}, \phi_{\mathcal{F}}\right)\right). \quad (9)$$

As our depolarized holography generates different images based on the polarization state of light, we obtain a polarization-dependent dataset of captured amplitudes for model training. The dataset consists of 1,600 phase patterns generated using stochastic gradient descent, and additional 400 phase patterns obtained through the alternating direction method of multipliers method [Choi et al. 2021]. Among the phase patterns generated using the stochastic gradient descent method, 800 phases were optimized with 2D images as the targets, while the remaining 800 phases were optimized using incoherent focal stacks derived from 2D images. In total, 2,000 phase patterns are used for training each channel. During the dataset generation, we randomized learning rates, propagation distances, and ranges of initial random phase distribution. We capture the intensity of holograms in 7 depth planes, encompassing 4 cases for a single phase pattern: without a metasurface, and vertical, horizontal, and diagonal polarization with the metasurface. This enables the model to learn the polarization-dependent phase and amplitude of the metasurface through CITL training, along with other parameters of the propagation model. The model is trained for 10 epochs with learning rate of $5e^{-4}$ for each channel. The model training takes approximately 6 hours for each channel on an NVIDIA RTX A6000. Additional details and analysis regarding the CITL model calibration and the results of trained parameters are provided in the Supplementary Material.

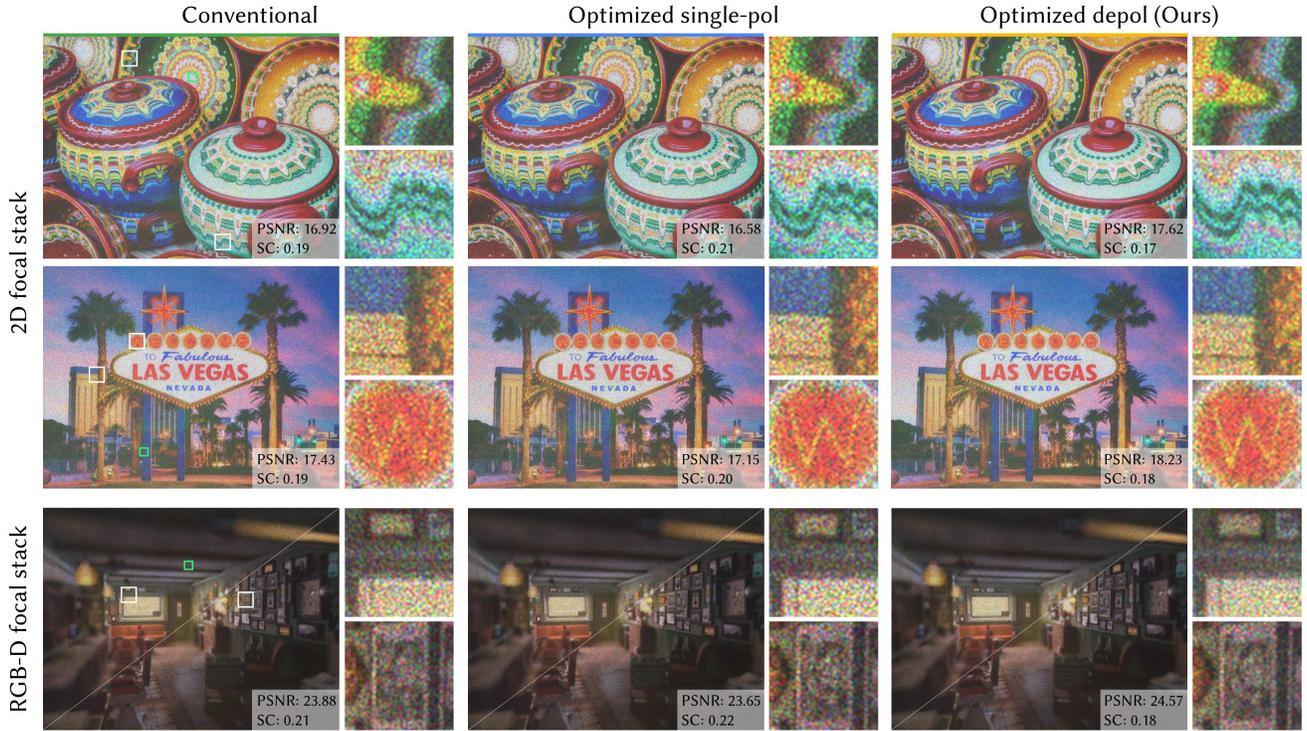

Fig. 9. The captured images of focal stack holograms, where the target focal stacks are generated from 2D images for the first and second rows, and from RGB-D data for the third row. In the third row, the upper left sub-images display the image focused at the far plane, while the lower right sub-images show the image focused at the near plane. The three columns, from left to right, correspond to the holographic display without a metasurface (*conventional*), with the optimized metasurface using a single polarization channel (*optimized single-pol*), and with the optimized metasurface using the two orthogonal polarization channels (*optimized depol*). Among these three cases, *optimized depol* exhibits the best image quality, with a smoother speckle intensity pattern and clearer details of the image. The PSNR and the speckle contrast values are provided in the bottom right of each image. In the actual experiments, the presence of unwanted DC noise decreases the variability in PSNR between different conditions when compared to the reconstructed results. The green boxes indicate the specified area at which the speckle contrast is calculated. Source images credit to Mila Drumeva (first row), Sean Pavone (second row), and Blender Animation Studio (third row).

### 5.2 Experimental results

Figure 9 shows the experimentally captured image of the focal stack holograms in our benchtop prototype setup. The SLM phase patterns are optimized with the CITL-calibrated model with the incoherent focal stacks derived from 2D images or RGB-D data. The *random depol* case is excluded from the experiment due to the obvious disadvantages over the *optimized depol* observed in the simulation and is vulnerable to alignment and fabrication errors. The comparison between the *conventional* and *optimized depol* demonstrate that the inclusion of the metasurface results in the improved image quality of focal stack holograms. The grainy speckle pattern in the *conventional* case becomes smoother in the *optimized depol* case, resulting in enhanced visibility of image details. The enlargements of each image show that *optimized depol* case shows lower intensity fluctuation and reduced grainy patterns than the *conventional* one. The speckle pattern is high-frequency noise and obscures the mid-high frequency areas of the image with the *conventional* case. In contrast to that, the *optimized depol* enables the image details more visible by reducing the speckle noise. Also, the captured images with the *optimized single-pol* are even noisier than those without the metasurface, demonstrating severe speckle noise. As the polarization-multiplexing functionality enables the incoherent superposition of two polarization states, the integration of two complementary images results in the best image quality observed in the *optimized depol*. These experimental results align with the simulation, exhibiting the competence of our depolarized holography in image quality.

We note here that the image quality of our captured results falls below that of the state-of-the-art research, primarily due to the challenges in generating incoherent focal stacks using a single SLM phase pattern. It is well-known that achieving incoherent focal stack optimization with a coherent single frame phase pattern is challenging due to the presence of speckle noise and the problem being over-constrained. As a result, previous works have utilized time-multiplexed frames to generate incoherent focal stacks [Choi et al. 2022; Lee et al. 2022] or introduced specific constraints to facilitate the optimization process [Choi et al. 2021; Yang et al. 2022]. The performance of our method can be further improved by integrating it with other speckle reduction methods, as discussed in Section 6.

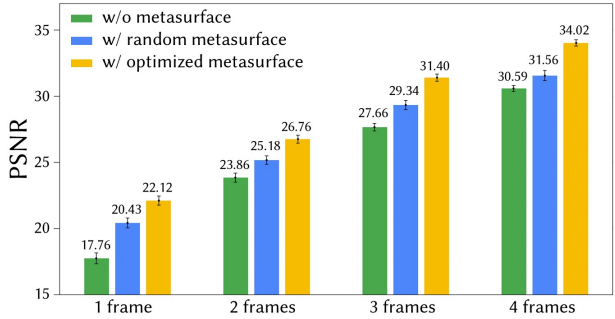

Fig. 10. The average PSNR of the time-multiplexed holographic display, combined with our depolarized holography enabled by metasurfaces. The PSNR values are obtained from 30 images of the DIV2K validation set, and the error bars indicate the standard error. The metasurfaces improve PSNR regardless of the number of time-multiplexed frames, demonstrating the advantage of our method under identical number of frames. Image quality results when our study is combined with a time-multiplexing scheme.

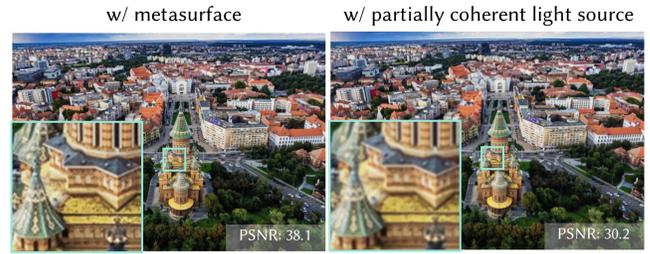

Fig. 11. Image quality comparison between our depolarized holography with the optimized metasurface (left) and the holographic display with a partially coherent light source (right). The measured PSNR of the reconstructed 2D image is provided at the right bottom of each image and the enlargements are additionally provided for visibility. Note that the CGHs are supervised with a sole 2D intensity profile, not with a focal stack. Source image credits to Salomia Oana Irina.

## 6 DISCUSSION

In this work, we introduce a depolarized holography enabled by the polarization-multiplexing metasurface to leverage the polarization channel of the holographic display. This novel approach allows for exploiting the mutual incoherence between orthogonal polarization states as a new degree of freedom for CGH optimization and speckle suppression. To this end, we present a joint optimization pipeline for co-designing the metasurface and the SLM phase patterns. Simulation results demonstrate that our scheme is superior to the conventional case, while the metasurface optimization further improves the image quality. Furthermore, we fabricate the optimized polarization-multiplexing metasurface and validate the proposed method using a display prototype. The experimental results align with the simulation and outperform conventional holographic displays.

### 6.1 Comparison with other speckle reduction methods

*Time-multiplexing scheme.* The proposed method stands independently and is compatible with other speckle reduction methods, demonstrating superior performance under identical conditions when combined. We firmly believe that incorporating an additional degree of freedom, brought by polarization-multiplexing metasurface, holds great advantages since other existing methods have their inherent limitations. For example, recently introduced time-multiplexed holographic displays report state-of-the-art results with speckleless, photorealistic images [Choi et al. 2022; Lee et al. 2022].

However, these displays necessitate high-speed SLMs, such as ferroelectric LC-based or MEMS-based SLMs, capable of rendering multiple frames within the flicker threshold of the human eye (50 Hz). The limited bit depth of SLM results in a decline in contrast, and the computation time or memory capacity escalates proportionally with the number of frames.

Note that the state-of-the-art LC technologies typically provide refresh rates up to approximately 400 Hz for 8-bit modulation [Zou et al. 2021], but it is important to highlight that these specific models may not be widely accessible on the commercial market. In this aspect, adopting our method with time-multiplexed holographic displays helps improve the image quality in a limited frame to utilize. Figure 10 shows simulation results of PSNR and speckle contrast estimated in time-multiplexed holographic displays integrated with the polarization-mutliplexing metasurface. The image quality metrics are measured from focal stack holograms generated from 30 images of DIV2K dataset [Agustsson and Timofte 2017]. Our method offers distinct advantages through its flexible integration with conventional time-multiplexing techniques and its ability to achieve competitive performance in terms of improved image metrics.

*Partially coherent light source.* A comparison between a depolarized holography and a holographic display using a partially coherent light source provides valuable insights into their performance. Figure 11 shows the simulation results of 2D holograms realized with holographic displays with a partially coherent source and our depolarized holography. To simulate the partially coherent light source, we assume a setup consisting of a collimating lens with a focal length of 200 mm and a light source with a square aperture of 100 $\mu m$ width. The wavelength spectrum of the light source is modeled to follow a Gaussian distribution with a standard deviation of 1 nm. The depolarized holography produces a sharp image with a higher PSNR compared to the image reconstructed using the partially coherent source. This observation underscores the image quality improvement achieved through our depolarized holography approach.

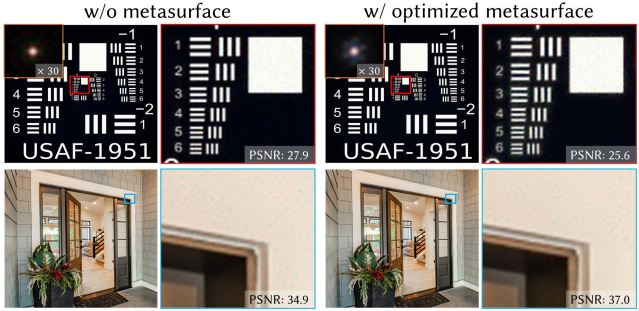

Fig. 12. Simulation results with various 2D images realized using conventional holographic displays w/o metasurface (left) and our depolarized holography w/ optimized metasurface (right). The insets in the top left corner of the first row show point spread functions magnified 30 times. The first row shows the 1951 USAF resolution target as a representative of binary 2D images, while the second row demonstrates natural 2D images. A section is cropped and enlarged with the PSNR provided at the bottom right. Source image credits to Sheila Say.

### 6.2 Challenges and future works

*Trade-offs between speckle reduction and image contrast.* In our work, the polarization-dependent phase modulation of the metasurface does not allow for the complete independent modulation of the two orthogonal polarization states, introducing noise under certain conditions. For example, we can think of the point spread function. In conventional holographic displays, a lens phase function provides a straightforward solution that generates a single point in space. However, with our depolarized holography, the metasurface alters the phase profile away from the lens phase function, which results in leakage around the focal spot. Furthermore, since the image reconstruction relies on averaging the images of orthogonal polarization states, a closed-form solution to generate a single point is elusive. The insets in the top-left corner of Fig. 12 visualize how the point spread function becomes blurred when using our polarization-multiplexing metasurface.

*Scene-specific holographic realization.* The optimization of the metasurface involves a training procedure with a set of natural images, which can introduce scene-specific limitations to our work. In certain exceptional cases, noise may undesirably appear. This is particularly true when reconstructing images with binary intensity distribution. The first row in Fig. 12 demonstrates the reconstructed binary 2D images realized by the conventional holographic display without metasurface, and our depolarized holography with the optimized metasurface. Here, we provide the reconstructed images of a resolution target, which can represent the aforementioned exceptional case. The reconstructed image with our method suffers from a slight degradation in the image quality evaluated with PSNR. However, the second row of the figure showcases that our method is beneficial in 2D natural images, even though not included in the metasurface optimization process.

*Optimization algorithms.* There is room for further improvement in the optimization algorithms used for both metasurfaces and SLM

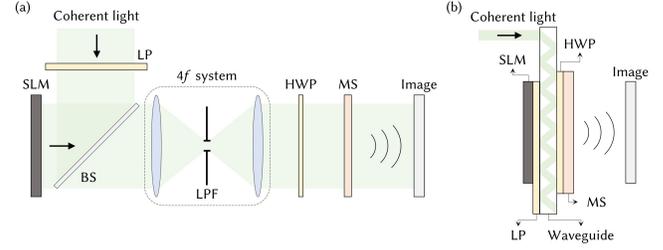

Fig. 13. (left) Our benchtop prototype utilizes a $4f$ system to relay the SLM and exactly matches the metasurface with it. However, the position of the metasurface can be adjusted freely, only if it can be incorporated in the joint optimization pipeline. (right) The potential compact display scheme utilizes waveguides, thus placing the SLM and metasurface on opposite sides of it. This design significantly reduces the form factor and realizes very lightweight holographic display devices.

phase patterns. While our current approach optimizes the metasurface in a per-pixel manner, previous research has reported that using a basis for designing optical elements can help avoid the pitfalls of local minima [Chang and Wetzstein 2019; Sun et al. 2020a]. Additionally, the SLM phase patterns are generated with the iterative stochastic gradient method, which currently takes several tens of seconds to converge. To enable real-time applications, it may be beneficial to explore the use of a deep neural network for CGH optimization in combination with the optimization pipeline, as demonstrated in many end-to-end cameras [Shi et al. 2022; Sun et al. 2020b]. Lastly, while we optimize the metasurface with an ideal wave propagation model and use CITL-calibration as a fine-tuning step, there is potential for further improvement by directly optimizing the metasurface with a CITL-calibrated model. This approach holds promise for enhancing the performance of the metasurface in real-world applications.

*Fabrication cost of metasurface.* In this work, the metasurface is fabricated utilizing electron beam lithography equipment. This equipment is capable of producing nanopatterns with up to tens of nanometer resolution, but typically have lower throughput due to the time-consuming nature of scanning the electron beam across the substrate, and thus, expensive cost. To solve these issues, Lee et al. [2018] and Yoon et al. [2020] show the potential for low-cost mass production of large-area metasurfaces using a method called nano-imprinting, in which only the master mold is produced by electron beam lithography and then replicas are printed in large quantities. Utilizing stepper photolithography being widely used in semiconductor fabrication efficiently produces large-area metasurfaces [Leitis et al. 2021; Park et al. 2019; She et al. 2018]. Most recently, Kim et al. [2023] achieves the extreme practicality of metasurface fabrication by combining photolithography with wafer-scale nanoimprinting. Using these approaches, large-area metasurfaces can be mass-produced at low cost, which has great potential for industrial applications, including holographic displays.

*System form factor.* In our benchtop prototype, we utilize a $4f$ system to relay the SLM directly onto the metasurface so that the two devices are in the same plane for experimental convenience (Figure 13(a)). However, it is not necessary for the metasurface to be precisely located in the relayed SLM plane. As long as the position of the metasurface in the optical path is known, it can be incorporated into the joint optimization pipeline, regardless of its location. Therefore, to further miniaturize the device in a practical manner, an alternative way is to position the SLM and the metasurface on opposite sides of the waveguide [Kim et al. 2022a; Maimone and Wang 2020]. This configuration eliminates the requirement for the $4f$ relay optics, which significantly contributes to the overall form factor of the current system. In the proposed design, the metasurface and SLM are separated by the thickness of the waveguide (Fig. 13(b)). The identical optimization pipeline used in this work can be applied to realize a thin and lightweight holographic display platform, resembling the form factor of sunglasses [Lee et al. 2018]. The compact and lightweight nature of the metasurface makes it an optimized optical element for such wearable devices. While the system form factor issue is beyond the scope of this study, it presents an interesting and meaningful topic for future research. There are also several examples of combining metasurfaces with liquid crystals. By integrating the metasurface with the SLM in the fabrication process [Badloe et al. 2022; Li et al. 2019], not only the system form factor but also the alignment issue between the SLM and the metasurface can be solved.

*Human factors.* In our work, we focus on speckle reduction to improve the image quality; as a result, the étendue of our system is identical to conventional holographic displays. As a narrow étendue limits the field of view and the eyebox of the display, étendue expansion is widely recognized as a core challenge in achieving practical applications for holographic displays. Efforts have been made to expand the étendue through various optical elements such as binary masks [Kuo et al. 2020], diffractive optical elements [Baek et al. 2021b], and lens arrays [Chae et al. 2023]. While we do not address the étendue expansion in this work, exploring the application of metasurfaces for the étendue expansion appears promising. The complex modulation and polarization-multiplexing capabilities of metasurfaces have the potential to further enhance the quality of holographic displays with étendue expansion.

In addition, throughout the paper, we assume that all the light from the SLM is observed. However, the eyebox is sampled by the ocular pupil in practical viewing scenarios, and this leads to changes in the perceived image and speckle pattern of the reconstructed image [Chakravarthula et al. 2022a, 2021]. Although we anticipate that our method may remain effective with pupil sampling, optimizing the metasurface incorporating the pupil sampling effect could be an interesting future work.

## 7 CONCLUSION

Prompted by state-of-the-art CGH algorithms, holographic displays have made significant strides in achieving photorealistic images. However, the physical aspects of holographic displays, which define fundamental limitations, have often been overlooked. In this study, we demonstrate the polarization-multiplexed holographic display using a novel optical platform called metasurface, and expand the degree of freedom in CGH optimization. We believe that our work serves as a milestone in a new approach to leverage the unprecedented optical functionality of nano-optics to address unsolved challenges in holographic displays as well as various conventional optical systems.

## REFERENCES

Eirikur Agustsson and Radu Timofte. 2017. NTIRE 2017 Challenge on Single Image Super-Resolution: Dataset and Study. In *The IEEE Conference on Computer Vision and Pattern Recognition (CVPR) Workshops*.

Amir Arbabi, Yu Horie, Mahmood Bagheri, and Andrei Faraon. 2015a. Dielectric metasurfaces for complete control of phase and polarization with subwavelength spatial resolution and high transmission. *Nature Nanotechnology* 10, 11 (2015), 937–943. arXiv:1411.1494

Amir Arbabi, Yu Horie, Alexander J. Ball, Mahmood Bagheri, and Andrei Faraon. 2015b. Subwavelength-thick lenses with high numerical apertures and large efficiency based on high-contrast transmitarrays. *Nature Communications* 6, 1 (2015), 7069.

Ehsan Arbabi, Jiaqi Li, Romanus J. Hutchins, Seyedeh Mahsa Kamali, Amir Arbabi, Yu Horie, Pol Van Dorpe, Viviana Gradinaru, Daniel A. Wagenaar, and Andrei Faraon. 2018. Two-Photon Microscopy with a Double-Wavelength Metasurface Objective Lens. *Nano Letters* 18, 8 (2018), 4943–4948.

Trevon Badloe, Joohoon Kim, Inki Kim, Won-Sik Kim, Wook Sung Kim, Young-Ki Kim, and Junsuk Rho. 2022. Liquid crystal-powered Mie resonators for electrically tunable photorealistic color gradients and dark blacks. *Light: Science & Applications* 11, 1 (2022), 118.

Seung-Hwan Baek and Felix Heide. 2021. Polarimetric spatio-temporal light transport probing. *ACM Transactions on Graphics* 40, 6 (2021), 1–18.

Seung-Hwan Baek, Hayato Ikoma, Daniel S. Jeon, Yuqi Li, Wolfgang Heidrich, Gordon Wetzstein, and Min H. Kim. 2021a. Single-Shot Hyperspectral-Depth Imaging With Learned Diffractive Optics. In *Proceedings of the IEEE/CVF International Conference on Computer Vision (ICCV)*. 2651–2660.

Seung-Hwan Baek, Ethan Tseng, Andrew Maimone, Nathan Matsuda, Grace Kuo, Qiang Fu, Wolfgang Heidrich, Douglas Lanman, and Felix Heide. 2021b. Neural \'{E}tendue Expander for Ultra-Wide-Angle High-Fidelity Holographic Display. *arXiv* (2021). arXiv:2109.08123

Vittorio Bianco, Pasquale Memmolo, Marco Leo, Silvio Montresor, Cosimo Distante, Melania Paturzo, Pascal Picart, Bahram Javidi, and Pietro Ferraro. 2018. Strategies for reducing speckle noise in digital holography. *Light: Science & Applications* 7, 1 (2018), 48.

David Blinder, Maksymilian Chlipala, Tomasz Kozacki, and Peter Schelkens. 2021. Photorealistic computer generated holography with global illumination and path tracing. *Optics Letters* 46, 9 (2021), 2188.

Minseok Chae, Kiseung Bang, Mingheon Yoo, and Yoonchan Jeong. 2023. Étendue Expansion in Holographic Near Eye Displays through Sparse Eye-Box Generation Using Lens Array Eyepiece. *ACM Transactions on Graphics* 42, 4 (2023).

Ayan Chakrabarti. 2016. Learning Sensor Multiplexing Design through Back-propagation. In *Advances in Neural Information Processing Systems*, D. Lee, M. Sugiyama, U. Luxburg, I. Guyon, and R. Garnett (Eds.), Vol. 29. Curran Associates, Inc. https://proceedings.neurips.cc/paper_files/paper/2016/file/aa486f25175cbdc3854151288a645c19-Paper.pdf

Praneeth Chakravarthula, Seung-Hwan Baek, Florian Schiffers, Ethan Tseng, Grace Kuo, Andrew Maimone, Nathan Matsuda, Oliver Cossairt, Douglas Lanman, and Felix Heide. 2022a. Pupil-Aware Holography. *ACM Transactions on Graphics* 41, 6, Article 212 (2022).

Praneeth Chakravarthula, Yifan Peng, Joel Kollin, Henry Fuchs, and Felix Heide. 2019. Wirtinger Holography for Near-Eye Displays. 38, 6, Article 213 (2019), 13 pages.

Praneeth Chakravarthula, Ethan Tseng, Henry Fuchs, and Felix Heide. 2022b. Hogel-Free Holography. *ACM Trans. Graph.* 41, 5, Article 178 (oct 2022), 16 pages.

Praneeth Chakravarthula, Ethan Tseng, Tarun Srivastava, Henry Fuchs, and Felix Heide. 2020. Learned Hardware-in-the-Loop Phase Retrieval for Holographic near-Eye Displays. *ACM Trans. Graph.* 39, 6, Article 186 (2020), 18 pages.

Praneeth Chakravarthula, Zhan Zhang, Okan Tursun, Piotr Didyk, Qi Sun, and Henry Fuchs. 2021. Gaze-Contingent Retinal Speckle Suppression for Perceptually-Matched Foveated Holographic Displays. *IEEE Transactions on Visualization and Computer Graphics* 27, 11 (2021), 4194–4203. https://doi.org/10.1109/TVCG.2021.3106433

Chenliang Chang, Kiseung Bang, Gordon Wetzstein, Byoungho Lee, and Liang Gao. 2020. Toward the next-generation VR/AR optics: a review of holographic near-eye displays from a human-centric perspective. *Optica* 7, 11 (2020), 1563–1578.

Julie Chang and Gordon Wetzstein. 2019. Deep Optics for Monocular Depth Estimation and 3D Object Detection. In *Proc. IEEE ICCV*.

Hou-Tong Chen, Antoinette J Taylor, and Nanfang Yu. 2016. A review of metasurfaces: physics and applications. *Reports on Progress in Physics* 79, 7 (2016), 076401. arXiv:1605.07672

Suyeon Choi, Manu Gopakumar, Yifan Peng, Jonghyun Kim, Matthew O'Toole, and Gordon Wetzstein. 2022. Time-Multiplexed Neural Holography: A Flexible Framework for Holographic Near-Eye Displays with Fast Heavily-Quantized Spatial Light Modulators. In *ACM SIGGRAPH 2022 Conference Proceedings* (Vancouver, BC, Canada) *(SIGGRAPH '22)*. Association for Computing Machinery, New York, NY, USA, Article 32, 9 pages.

Suyeon Choi, Manu Gopakumar, Yifan Peng, Jonghyun Kim, and Gordon Wetzstein. 2021. Neural 3D Holography: Learning Accurate Wave Propagation Models for 3D Holographic Virtual and Augmented Reality Displays. *ACM Trans. Graph.* 40, 6, Article 240 (2021), 12 pages.

Haejun Chung and Owen D Miller. 2020. High-NA achromatic metalenses by inverse design. *Optics Express* 28, 5 (2020), 6945. arXiv:1905.09213

Manuel Decker, Isabelle Staude, Matthias Falkner, Jason Dominguez, Dragomir N. Neshev, Igal Brener, Thomas Pertsch, and Yuri S. Kivshar. 2015. High-Efficiency Dielectric Huygens' Surfaces. *Advanced Optical Materials* 3, 6 (2015), 813–820. arXiv:1405.5038

Yuanbo Deng and Daping Chu. 2017. Coherence properties of different light sources and their effect on the image sharpness and speckle of holographic displays. *Scientific Reports* 7, 1 (2017), 1–12.

Xiong Dun, Hayato Ikoma, Gordon Wetzstein, Zhanshan Wang, Xinbin Cheng, and Yifan Peng. 2020. Learned rotationally symmetric diffractive achromat for full-spectrum computational imaging. *Optica* 7, 8 (Aug 2020), 913–922. https://doi.org/10.1364/OPTICA.394413

James R Fienup. 1982. Phase retrieval algorithms: a comparison. *Applied Optics* 21, 15 (1982), 2758–2769.

Ralph W Gerchberg. 1972. A practical algorithm for the determination of phase from image and diffraction plane pictures. *Optik* 35 (1972), 237–246.

Joseph W Goodman. 2005. *Introduction to Fourier optics*. Roberts and Company Publishers.

Joseph W Goodman. 2007. *Speckle phenomena in optics: theory and applications*. Roberts and Company Publishers.

Jinying Guo, Teng Wang, Baogang Quan, Huan Zhao, Changzhi Gu, Junjie Li, Xinke Wang, Guohai Situ, and Yan Zhang. 2019. Polarization multiplexing for double images display. *Opto-Electronic Advances* 2, 11 (7 2019), 1. https://www.researching.cn/articles/OJ7117d3f213bd5850

Harel Haim, Shay Elmalem, Raja Giryes, Alex M. Bronstein, and Emanuel Marom. 2018. Depth Estimation From a Single Image Using Deep Learned Phase Coded Mask. *IEEE Transactions on Computational Imaging* 4, 3 (2018), 298–310. https://doi.org/10.1109/TCI.2018.2849326

Dean S Hazineh, Soon Wei Daniel Lim, Zhujun Shi, Federico Capasso, Todd Zickler, and Qi Guo. 2022. D-Flat: A Differentiable Flat-Optics Framework for End-to-End Metasurface Visual Sensor Design. *arXiv* (2022).

Hui-Hsin Hsiao, Cheng Hung Chu, and Din Ping Tsai. 2017. Fundamentals and Applications of Metasurfaces. *Small Methods* 1, 4 (2017), 1600064.

Lingling Huang, Xianzhong Chen, Holger Mühlenbernd, Hao Zhang, Shumei Chen, Benfeng Bai, Qiaofeng Tan, Guofan Jin, Kok-Wai Cheah, Cheng-Wei Qiu, Jensen Li, Thomas Zentgraf, and Shuang Zhang. 2013. Three-dimensional optical holography using a plasmonic metasurface. *Nature Communications* 4, 1 (2013), 2808.

Inseung Hwang, Daniel S Jeon, Adolfo Muñoz, Diego Gutierrez, Xin Tong, and Min H Kim. 2022. Sparse ellipsometry: portable acquisition of polarimetric SVBRDF and shape with unstructured flash photography. *ACM Transactions on Graphics* 41, 4 (2022), 1–14. arXiv:2207.04236

Michael Iliadis, Leonidas Spinoulas, and Aggelos K. Katsaggelos. 2020. DeepBinaryMask: Learning a binary mask for video compressive sensing. *Digital Signal Processing* 96 (2020), 102591. https://doi.org/10.1016/j.dsp.2019.102591

Changwon Jang, Kiseung Bang, Minseok Chae, Byoungho Lee, and Douglas Lanman. 2022. Waveguide Holography: Towards True 3D Holographic Glasses.

Junhyeok Jang, Gun-Yeal Lee, Jangwoon Sung, and Byoungho Lee. 2021. Independent Multichannel Wavefront Modulation for Angle Multiplexed Meta-Holograms. *Advanced Optical Materials* 9, 17 (2021), 2100678.

Seyedeh Mahsa Kamali, Ehsan Arbabi, Amir Arbabi, Yu Horie, MohammadSadegh Faraji-Dana, and Andrei Faraon. 2017. Angle-Multiplexed Metasurfaces: Encoding Independent Wavefronts in a Single Metasurface under Different Illumination Angles. 7, 4 (2017), 041056.

Mohammadreza Khorasaninejad, Wei Ting Chen, Robert C. Devlin, Jaewon Oh, Alexander Y. Zhu, and Federico Capasso. 2016. Metalenses at visible wavelengths: Diffraction-limited focusing and subwavelength resolution imaging. *Science* 352, 6290 (2016), 1190–1194.

Changhyun Kim, Sun-Je Kim, and Byoungho Lee. 2020. Doublet metalens design for high numerical aperture and simultaneous correction of chromatic and monochromatic aberrations. 28, 12 (2020), 18059.

Changhyun Kim and Byoungho Lee. 2023. TORCWA: GPU-accelerated Fourier modal method and gradient-based optimization for metasurface design. *Computer Physics Communications* 282 (2023), 108552.

Changil Kim, Henning Zimmer, Yael Pritch, Alexander Sorkine-Hornung, and Markus H Gross. 2013. Scene reconstruction from high spatio-angular resolution light fields. *ACM Trans. Graph.* 32, 4 (2013), 73–1.

Dongyeon Kim, Seung-Woo Nam, Kiseung Bang, Byounghyo Lee, Seungjae Lee, Youngmo Jeong, Jong-Mo Seo, and Byoungho Lee. 2021b. Vision-correcting holographic display: evaluation of aberration correcting hologram. *Biomedical Optics Express* 12, 8 (2021), 5179–5195.

Dongyeon Kim, Seung-Woo Nam, Byounghyo Lee, Jong-Mo Seo, and Byoungho Lee. 2022b. Accommodative Holography: Improving Accommodation Response for Perceptually Realistic Holographic Displays. *ACM Trans. Graph.* 41, 4, Article 111 (jul 2022), 15 pages.

Jonghyun Kim, Manu Gopakumar, Suyeon Choi, Yifan Peng, Ward Lopes, and Gordon Wetzstein. 2022a. Holographic Glasses for Virtual Reality. In *ACM SIGGRAPH 2022 Conference Proceedings* (Vancouver, BC, Canada) *(SIGGRAPH '22)*. Association for Computing Machinery, New York, NY, USA, Article 33, 9 pages.

Joohoon Kim, Junhwa Seong, Wonjoong Kim, Gun-Yeal Lee, Seokwoo Kim, Hongyoon Kim, Seong-Won Moon, Dong Kyo Oh, Younghwan Yang, Jeonghoon Park, Jaehyuck Jang, Yeseul Kim, Minsu Jeong, Chanwoong Park, Hojung Choi, Gyoseon Jeon, Kyung-il Lee, Dong Hyun Yoon, Namkyoo Park, Byoungho Lee, Heon Lee, and Junsuk Rho. 2023. Scalable manufacturing of high-index atomic layer–polymer hybrid metasurfaces for metaphotonics in the visible. *Nature Materials* 22 (2023), 474–481.

Youngjin Kim, Gun-Yeal Lee, Jangwoon Sung, Junhyeok Jang, and Byoungho Lee. 2021a. Spiral Metalens for Phase Contrast Imaging. *Advanced Functional Materials* (2021).

Tomasz Kozacki and Maksymilian Chlipala. 2016. Color holographic display with white light LED source and single phase only SLM. *Optics Express* 24, 3 (2016), 2189–2199.

Grace Kuo, Laura Waller, Ren Ng, and Andrew Maimone. 2020. High Resolution éTendue Expansion for Holographic Displays. *ACM Trans. Graph.* 39, 4, Article 66 (2020), 14 pages.

Byounghyo Lee, Dongyeon Kim, Seungjae Lee, Chun Chen, and Byoungho Lee. 2022. High-contrast, speckle-free, true 3D holography via binary CGH optimization. *Scientific Reports* 12, 1 (18 Feb 2022), 2811.

Gun-Yeal Lee, Jong-Young Hong, SoonHyoung Hwang, Seokil Moon, Hyeokjung Kang, Sohee Jeon, Hwi Kim, Jun-Ho Jeong, and Byoungho Lee. 2018. Metasurface eyepiece for augmented reality. *Nature Communications* 9, 1 (2018), 4562.

Gun-Yeal Lee, Jangwoon Sung, and Byoungho Lee. 2019. Recent advances in metasurface hologram technologies (Invited paper). *ETRI Journal* 41, 1 (2019), 10–22.

Gun-Yeal Lee, Gwanho Yoon, Seung-Yeol Lee, Hansik Yun, Jaebum Cho, Kyookeun Lee, Hwi Kim, Junsuk Rho, and Byoungho Lee. 2017. Complete amplitude and phase control of light using broadband holographic metasurfaces. *Nanoscale* 10, 9 (2017), 4237–4245. arXiv:1706.09632

Seungjae Lee, Dongyeon Kim, Seung-Woo Nam, Byounghyo Lee, Jaebum Cho, and Byoungho Lee. 2020. Light source optimization for partially coherent holographic displays with consideration of speckle contrast, resolution, and depth of field. *Scientific Reports* 10, 1 (2020), 1–12.

Aleksandrs Leitis, Andreas Tittl, Mingkai Liu, Bang Hyun Lee, Man Bock Gu, Yuri S. Kivshar, and Hatice Altug. 2019. Angle-multiplexed all-dielectric metasurfaces for broadband molecular fingerprint retrieval. *Science Advances* 5, 5 (2019), eaaw2871. https://doi.org/10.1126/sciadv.aaw2871

Aleksandrs Leitis, Ming Lun Tseng, Aurelian John-Herpin, Yuri S. Kivshar, and Hatice Altug. 2021. Wafer-Scale Functional Metasurfaces for Mid-Infrared Photonics and Biosensing. *Advanced Materials* (2021), 2102232.

Gang Li, Dukho Lee, Youngmo Jeong, Jaebum Cho, and Byoungho Lee. 2016. Holographic display for see-through augmented reality using mirror-lens holographic optical element. *Optics Letters* 41, 11 (2016), 2486–2489.

Shi-Qiang Li, Xuewu Xu, Rasna Maruthiyodan Veetil, Vytautas Valuckas, Ramón Paniagua-Domínguez, and Arseniy I. Kuznetsov. 2019. Phase-only transmissive spatial light modulator based on tunable dielectric metasurface. *Science* 364, 6445 (2019), 1087–1090.

Zhaoyi Li, Peng Lin, Yao-Wei Huang, Joon-Suh Park, Wei Ting Chen, Zhujun Shi, Cheng-Wei Qiu, Ji-Xin Cheng, and Federico Capasso. 2021. Meta-optics achieves RGB-achromatic focusing for virtual reality. *Science Advances* 7, 5 (2021), eabe4458.

Zhaoyi Li, Raphaël Pestourie, Joon-Suh Park, Yao-Wei Huang, Steven G. Johnson, and Federico Capasso. 2022. Inverse design enables large-scale high-performance meta-optics reshaping virtual reality. *Nature Communications* 13, 1 (2022), 2409. arXiv:2104.09702

Dianmin Lin, Pengyu Fan, Erez Hasman, and Mark L. Brongersma. 2014. Dielectric gradient metasurface optical elements. *Science* 345, 6194 (2014), 298–302.

Andrew Maimone, Andreas Georgiou, and Joel S Kollin. 2017. Holographic near-eye displays for virtual and augmented reality. *ACM Transactions on Graphics (Tog)* 36, 4 (2017), 1–16.

Andrew Maimone and Junren Wang. 2020. Holographic Optics for Thin and Lightweight Virtual Reality. *ACM Trans. Graph.* 39, 4, Article 67 (aug 2020), 14 pages. https://doi.org/10.1145/3386569.3392416

Maksim Makarenko, Arturo Burguete-Lopez, Qizhou Wang, Fedor Getman, Silvio Giancola, Bernard Ghanem, and Andrea Fratalocchi. 2022. Real-time Hyperspectral Imaging in Hardware via Trained Metasurface Encoders. *arXiv* (2022).

Mahdad Mansouree, Hyounghan Kwon, Ehsan Arbabi, Andrew McClung, Andrei Faraon, and Amir Arbabi. 2020. Multifunctional 25D metastructures enabled by adjoint optimization. *Optica* 7, 1 (2020), 77.

Kyoji Matsushima and Tomoyoshi Shimobaba. 2009. Band-limited angular spectrum method for numerical simulation of free-space propagation in far and near fields. *Optics express* 17, 22 (2009), 19662–19673.

Christopher A. Metzler, Hayato Ikoma, Yifan Peng, and Gordon Wetzstein. 2020. Deep Optics for Single-Shot High-Dynamic-Range Imaging. In *Proceedings of the IEEE/CVF Conference on Computer Vision and Pattern Recognition (CVPR)*.

J. P. Balthasar Mueller, Noah A. Rubin, Robert C. Devlin, Benedikt Groever, and Federico Capasso. 2017. Metasurface Polarization Optics: Independent Phase Control of Arbitrary Orthogonal States of Polarization. *Physical Review Letters* 118, 11 (2017), 113901.

Seung-Woo Nam, Dongyeon Kim, and Byoungho Lee. 2022. Accelerating a spatially varying aberration correction of holographic displays with low-rank approximation. *Opt. Lett.* 47, 13 (Jul 2022), 3175–3178.

Seung-Woo Nam, Seokil Moon, Byounghyo Lee, Dongyeon Kim, Seungjae Lee, Chang-Kun Lee, and Byoungho Lee. 2020. Aberration-corrected full-color holographic augmented reality near-eye display using a Pancharatnam-Berry phase lens. *Optics Express* 28, 21 (2020), 30836.

Adam C. Overvig, Sajan Shrestha, Stephanie C. Malek, Ming Lu, Aaron Stein, Changxi Zheng, and Nanfang Yu. 2019. Dielectric metasurfaces for complete and independent control of the optical amplitude and phase. *Light: Science & Applications* 8, 1 (2019), 92.

Nitish Padmanaban, Yifan Peng, and Gordon Wetzstein. 2019. Holographic Near-Eye Displays Based on Overlap-Add Stereograms. *ACM Trans. Graph.* 38, 6, Article 214 (nov 2019), 13 pages.

Jae-Hyeung Park. 2017. Recent progress in computer-generated holography for three-dimensional scenes. *Journal of Information Display* 18, 1 (2017), 1–12.

Joon-Suh Park, Shuyan Zhang, Alan She, Wei Ting Chen, Peng Lin, Kerolos M. A. Yousef, Ji-Xin Cheng, and Federico Capasso. 2019. All-Glass, Large Metalens at Visible Wavelength Using Deep-Ultraviolet Projection Lithography. *Nano Letters* 19, 12 (2019), 8673–8682.

Yifan Peng, Suyeon Choi, Jonghyun Kim, and Gordon Wetzstein. 2021. Speckle-free holography with partially coherent light sources and camera-in-the-loop calibration. *Science Advances* 7, 46 (2021), eabg5040.

Yifan Peng, Suyeon Choi, Nitish Padmanaban, and Gordon Wetzstein. 2020. Neural Holography with Camera-in-the-Loop Training. *ACM Trans. Graph.* 39, 6, Article 185 (nov 2020), 14 pages.

Yifan Peng, Qilin Sun, Xiong Dun, Gordon Wetzstein, Wolfgang Heidrich, and Felix Heide. 2019. Learned Large Field-of-View Imaging with Thin-Plate Optics. 38, 6, Article 219 (nov 2019), 14 pages. https://doi.org/10.1145/3355089.3356526

Raphaël Pestourie, Carlos Pérez-Arancibia, Zin Lin, Wonseok Shin, Federico Capasso, and Steven G Johnson. 2018. Inverse design of large-area metasurfaces. *Optics Express* 26, 26 (2018), 33732. arXiv:1808.04215

Lu Rong, Wen Xiao, Feng Pan, Shuo Liu, and Rui Li. 2010. Speckle noise reduction in digital holography by use of multiple polarization holograms. *Chin. Opt. Lett.* 8, 7 (Jul 2010), 653–655. https://opg.optica.org/col/abstract.cfm?URI=col-8-7-653

Bernard Rous. 2008. The Enabling of Digital Libraries. *Digital Libraries* 12, 3, Article 5 (July 2008). To appear.

Noah A. Rubin, Gabriele D'Aversa, Paul Chevalier, Zhujun Shi, Wei Ting Chen, and Federico Capasso. 2019. Matrix Fourier optics enables a compact full-Stokes polarization camera. *Science* 365, 6448 (2019), eaax1839.

Noah A. Rubin, Aun Zaidi, Ahmed H. Dorrah, Zhujun Shi, and Federico Capasso. 2021. Jones matrix holography with metasurfaces. *Science Advances* 7, 33 (2021), eabg7488.

Alan She, Shuyan Zhang, Samuel Shian, David R Clarke, and Federico Capasso. 2018. Large area metalenses: design, characterization, and mass manufacturing. *Optics Express* 26, 2 (2018), 1573.

Liang Shi, Fu-Chung Huang, Ward Lopes, Wojciech Matusik, and David Luebke. 2017. Near-Eye Light Field Holographic Rendering with Spherical Waves for Wide Field of View Interactive 3D Computer Graphics. *ACM Trans. Graph.* 36, 6, Article 236 (nov 2017), 17 pages.

Liang Shi, Beichen Li, Changil Kim, Petr Kellnhofer, and Wojciech Matusik. 2021. Towards real-time photorealistic 3D holography with deep neural networks. *Nature* 591, 7849 (2021), 234–239.

Zheng Shi, Yuval Bahat, Seung-Hwan Baek, Qiang Fu, Hadi Amata, Xiao Li, Praneeth Chakravarthula, Wolfgang Heidrich, and Felix Heide. 2022. Seeing through Obstructions with Diffractive Cloaking. *ACM Trans. Graph.* 41, 4, Article 37 (jul 2022), 15 pages.

Zhujun Shi, Mohammadreza Khorasaninejad, Yao-Wei Huang, Charles Roques-Carmes, Alexander Y Zhu, Wei Ting Chen, Vyshakh Sanjeev, Zhao-Wei Ding, Michele Tamagnone, Kundan Chaudhary, Robert C Devlin, Cheng-Wei Qiu, and Federico Capasso. 2018. Single-Layer Metasurface with Controllable Multiwavelength Functions. *Nano Letters* 18, 4 (2018), 2420–2427.

Vincent Sitzmann, Steven Diamond, Yifan Peng, Xiong Dun, Stephen Boyd, Wolfgang Heidrich, Felix Heide, and Gordon Wetzstein. 2018. End-to-End Optimization of Optics and Image Processing for Achromatic Extended Depth of Field and Super-Resolution Imaging. *ACM Trans. Graph.* 37, 4, Article 114 (jul 2018), 13 pages. https://doi.org/10.1145/3197517.3201333

Hend Sroor, Yao-Wei Huang, Bereneice Sephton, Darryl Naidoo, Adam Vallés, Vincent Ginis, Cheng-Wei Qiu, Antonio Ambrosio, Federico Capasso, and Andrew Forbes. 2020. High-purity orbital angular momentum states from a visible metasurface laser. 14, 8 (2020), 498–503.

Shlomi Steinberg and Ling-Qi Yan. 2021. A generic framework for physical light transport. *ACM Transactions on Graphics* 40, 4 (2021), 1–20.

Qilin Sun, Ethan Tseng, Qiang Fu, Wolfgang Heidrich, and Felix Heide. 2020a. Learning Rank-1 Diffractive Optics for Single-Shot High Dynamic Range Imaging. In *Proceedings of the IEEE/CVF Conference on Computer Vision and Pattern Recognition (CVPR)*.

Qilin Sun, Jian Zhang, Xiong Dun, Bernard Ghanem, Yifan Peng, and Wolfgang Heidrich. 2020b. End-to-End Learned, Optically Coded Super-Resolution SPAD Camera. *ACM Trans. Graph.* 39, 2, Article 9 (mar 2020), 14 pages.

Ethan Tseng, Shane Colburn, James Whitehead, Luocheng Huang, Seung-Hwan Baek, Arka Majumdar, and Felix Heide. 2021a. Neural nano-optics for high-quality thin lens imaging. *Nature Communications* 12, 1 (2021), 6493.

Ethan Tseng, Ali Mosleh, Fahim Mannan, Karl St-Arnaud, Avinash Sharma, Yifan Peng, Alexander Braun, Derek Nowrouzezahrai, Jean-François Lalonde, and Felix Heide. 2021b. Differentiable Compound Optics and Processing Pipeline Optimization for End-to-End Camera Design. 40, 2, Article 18 (jun 2021), 19 pages. https://doi.org/10.1145/3446791

Yicheng Wu, Vivek Boominathan, Huaijin Chen, Aswin Sankaranarayanan, and Ashok Veeraraghavan. 2019. PhaseCam3D — Learning Phase Masks for Passive Single View Depth Estimation. In *2019 IEEE International Conference on Computational Photography (ICCP)*. 1–12. https://doi.org/10.1109/ICCPHOT.2019.8747330

Daeho Yang, Wontaek Seo, Hyeonseung Yu, Sun Il Kim, Bongsu Shin, Chang-Kun Lee, Seokil Moon, Jungkwuen An, Jong-Young Hong, Geeyoung Sung, and Hong-Seok Lee. 2022. Diffraction-engineered holography: Beyond the depth representation limit of holographic displays. *Nature Communications* 13, 1 (2022), 6012.

Ozlem Yavas, Mikael Svedendahl, Paulina Dobosz, Vanesa Sanz, and Romain Quidant. 2017. On-a-chip Biosensing Based on All-Dielectric Nanoresonators. *Nano Letters* 17, 7 (2017), 4421–4426.

Han-Ju Yeom, Hee-Jae Kim, Seong-Bok Kim, HuiJun Zhang, BoNi Li, Yeong-Min Ji, Sang-Hoo Kim, and Jae-Hyeung Park. 2015. 3D holographic head mounted display using holographic optical elements with astigmatism aberration compensation. *Optics Express* 23, 25 (2015), 32025–32034.

Filiz Yesilkoy, Eduardo R. Arvelo, Yasaman Jahani, Mingkai Liu, Andreas Tittl, Volkan Cevher, Yuri Kivshar, and Hatice Altug. 2019. Ultrasensitive hyperspectral imaging and biodetection enabled by dielectric metasurfaces. *Nature Photonics* 13, 6 (2019), 390–396.

Gwanho Yoon, Kwan Kim, Daihong Huh, Heon Lee, and Junsuk Rho. 2020. Single-step manufacturing of hierarchical dielectric metalens in the visible. *Nature Communications* 11, 1 (2020), 2268.

Nanfang Yu, Patrice Genevet, Mikhail A. Kats, Francesco Aieta, Jean-Philippe Tetienne, Federico Capasso, and Zeno Gaburro. 2011. Light Propagation with Phase Discontinuities: Generalized Laws of Reflection and Refraction. *Science* 334, 6054 (2011), 333–337.

Jingzhao Zhang, Nicolas Pégard, Jingshan Zhong, Hillel Adesnik, and Laura Waller. 2017. 3D computer-generated holography by non-convex optimization. *Optica* 4, 10 (2017), 1306.

Guoxing Zheng, Holger Mühlenbernd, Mitchell Kenney, Guixin Li, Thomas Zentgraf, and Shuang Zhang. 2015. Metasurface holograms reaching 80% efficiency. *Nature Nanotechnology* 10, 4 (2015), 308–312.

You Zhou, Hanyu Zheng, Ivan I. Kravchenko, and Jason Valentine. 2020. Flat optics for image differentiation. *Nature Photonics* 14, 5 (2020), 316–323.

Junyu Zou, Qian Yang, En-Lin Hsiang, Haruki Ooiishi, Zhuo Yang, Kifumi Yoshidaya, and Shin-Tson Wu. 2021. Fast-Response Liquid Crystal for Spatial Light Modulator and LiDAR Applications. *Crystals* 11, 2, Article 93 (2021).

Xiujuan Zou, Youming Zhang, Ruoyu Lin, Guangxing Gong, Shuming Wang, Shining Zhu, and Zhenlin Wang. 2022. Pixel-level Bayer-type colour router based on metasurfaces. *Nature Communications* 13, 1 (2022), 3288.

# Depolarized Holography with Polarization-multiplexing Metasurface - Supplementary Material

SEUNG-WOO NAM* and YOUNGJIN KIM*, Seoul National University, Republic of Korea
DONGYEON KIM, Seoul National University, Republic of Korea
YOONCHAN JEONG, Seoul National University, Republic of Korea

Here we note some abbreviations frequently used throughout the supplementary text.

**RCWA** : Rigorous coupled-wave anaylsis

**SLM** : Spatial light modulator

## S1 ADDITIONAL DETAILS ON HARDWARE

### S1.1 Polarization-multiplexing metasurface

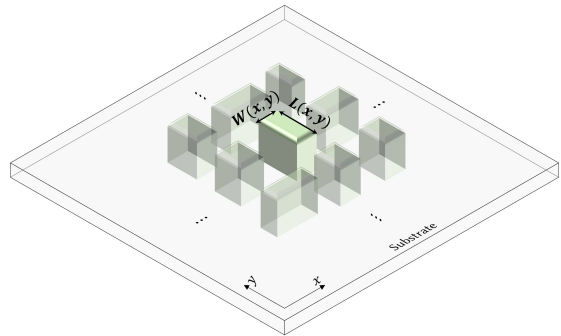

Fig. S1. Schematics of the polarization-multiplexing metasurface. Lateral dimensions along the $x$- and $y$- axes are written as $W(x, y)$ and $L(x, y)$, respectively. Each nanostructure can shift the phase of $x$- and $y$- polarized light by changing the $W$ and $L$.

*Principles of independent phase modulation for orthogonal linear polarization states.* Metasurfaces are two-dimensional arrays of nano-scatterer with a subwavelength period, as shown in Figure S1. Pixel-wise variation of geometric parameters, for instance, the length and width of rectangular-shaped nanorods can change quasi-independently the effective refractive indexes along the x- and y-axis, respectively. Thus, the phase shifts occur for each orthogonal linear polarization state. This optical behavior can be represented by the Jones matrix of linearly birefringent waveplate [Arbabi et al., 2015, Mueller et al., 2017].

$$\begin{bmatrix} e^{i\phi_x} & 0 \\ 0 & e^{i\phi_y} \end{bmatrix} \tag{S1}$$

---

*Phase modulation range of orthogonal linear polarization states.* In ideal case, the phase-shift of transmission coefficients for each orthogonal linear polarization states cover the whole $2\pi$ range theoretically, which means the complete independent modulation of orthogonal polarization-pair. As explained in the main text, however, the fabrication constraint or the kind of dielectric material we use might pose a hurdle for the complete independent phase modulation. Figure S2 shows the actual phase cover range along with practical issues; a low refractive index of the silicon nitride with a limited height of the nanorod. Each point in the figure represents the phase values for $t_{xx}$ and $t_{yy}$, respectively. Therefore, if it is possible to adjust the phase completely independently for two orthogonal polarizations, the points shown in the picture should be fully filled throughout the whole phase chart. As the wavelength of incident light increases, the range of possible values for the propagation phase scheme is reduced, assuming that the height of the nanorod is fixed. Thus, the phase modulation range at 638 nm wavelength shows much narrower than the case of 450 nm. The use of materials possessing higher refractive index such as titanium dioxide or amorphous silicon can be a simple solution to tackle with this problem. Also the realization of the sophisticated fabrication recipe enabling the higher aspect ratio is able to increase the phase-shift range, either.

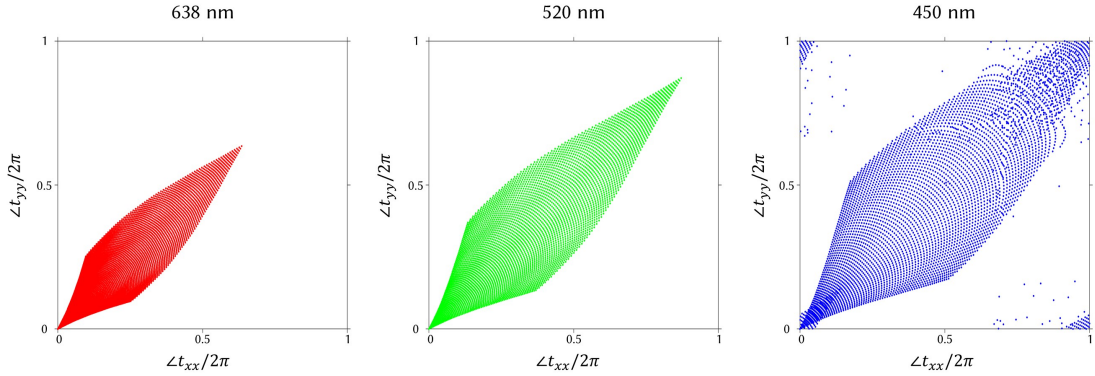

Fig. S2. Phase modulation range on the orthogonal linear polarization states at wavelengths of interest. $\angle t_{xx}$ and $\angle t_{yy}$ represent the phase shift of co-polarized transmission coefficient when it comes to the normal incidence of linearly polarized light. The phase shifts are normalized by $2\pi$.

*Proxy model fitting from the RCWA data.* Metasurface proxy model is designed from the pre-simulated transmittance of rectangular nanostructure calculated by the RCWA method. First, we have to specify several hyper-parameters that are decided by the experimental conditions. The pixel pitch of the metasurface is set to approximately 283 nm, determined under two considerations: a demagnification factor of the relay optics from SLM to metasurface and the suppression of the unwanted resonant phenomena inside the dielectric materials for smooth-fitting. Three wavelengths of the laser source are 450, 520, and 638 nm, respectively. The refractive index (n) and extinction coefficient (k) of the silicon nitride layer with a deposition thickness of 800 nm. Figure S3 shows the n, k values measured by spectroscopic ellipsometer (M2000D, Woollam). Second, given that the hyper-parameters are decided, we utilize the RCWA method to obtain transmittance libraries to be used for the proxy-model fitting. A total of six data sets on the combinations of the two phase shifts for each co-polarized transmission coefficient and the three different wavelengths, as a function of geometric parameters of the nanorod, which change from 80 to 220 nm with a 2 nm interval. For example, the phase shift of the co-polarized transmission coefficients is simulated by RCWA for every width and length value, when the x-polarized light is normally incident upon the nanostructure. Third, the discrete values of each library are fitted as a surface function using linear quadratic polynomials as explained in the main text. The phase shifts of transmitted light are also normalized by $2\pi$. We utilize the curve fitting toolbox from the commercial software, MATLAB. Using a linear-least-square method, the coefficients of polynomials can be obtained with a 95% confidence bound. Figure S4 shows the six proxy models against simulated values. Although we can see some outliers of the simulated data compared with the fitted functions, especially for the blue wavelength case, which is attributed to the resonant phenomena inside the dielectric materials, they are very sparse so we can neglect these exceptional points. Table 1 shows the equations and the coefficients of polynomials for all twelve proxy models.

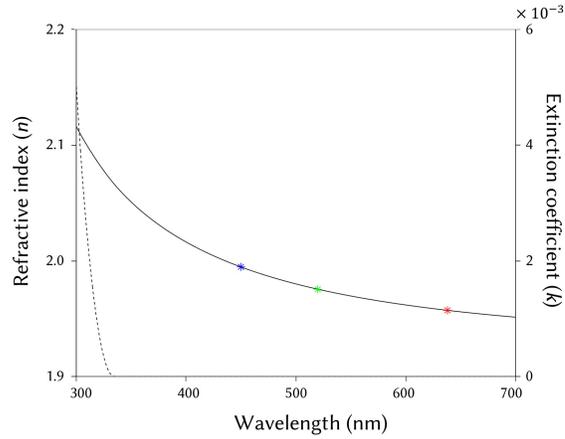

Fig. S3. Refractive index and extinction coefficient of the silicon nitride layer. The solid and dotted lines represent the refractive index (RI) and the extinction coefficient. The RI at wavelengths 450, 520, and 638 nm are marked with blue-, green-, and red-colored asterisks, which are 1.995, 1.976, and 1.957, respectively.

Table 1. Fitted coefficients of the linear quadratic polynomials. The coefficients of $c_{12}$, $c_{21}$, and $c_{22}$ are set to zeros. Superscripts 'r', 'g', and 'b' correspond to red, green, and blue. $t_{xx}$ defines the co-polarized transmission coefficient when the x-polarized light is normally incident upon the nanostructure.

| Physical entity | $c_{00}$ | $c_{10}$ | $c_{01}$ | $c_{20}$ | $c_{11}$ | $c_{02}$ |
|---|---|---|---|---|---|---|
| $\phi_{xx}^r$ | -0.0946 | -0.1171 | 0.06675 | 0.3065 | 1.204 | -0.2145 |
| $\phi_{xx}^g$ | -0.3072 | 0.3484 | 0.3064 | 0.05226 | 1.543 | -0.4258 |
| $\phi_{xx}^b$ | -0.7156 | 1.366 | 0.8043 | -0.5976 | 1.743 | -0.8002 |
| $\phi_{yy}^r$ | -0.09458 | 0.06663 | -0.1175 | -0.2144 | 1.204 | 0.3069 |
| $\phi_{yy}^g$ | -0.3072 | 0.3064 | 0.3486 | -0.4258 | 1.543 | 0.05215 |
| $\phi_{yy}^b$ | -0.7157 | 0.8048 | 1.365 | -0.8004 | 1.742 | -0.5967 |

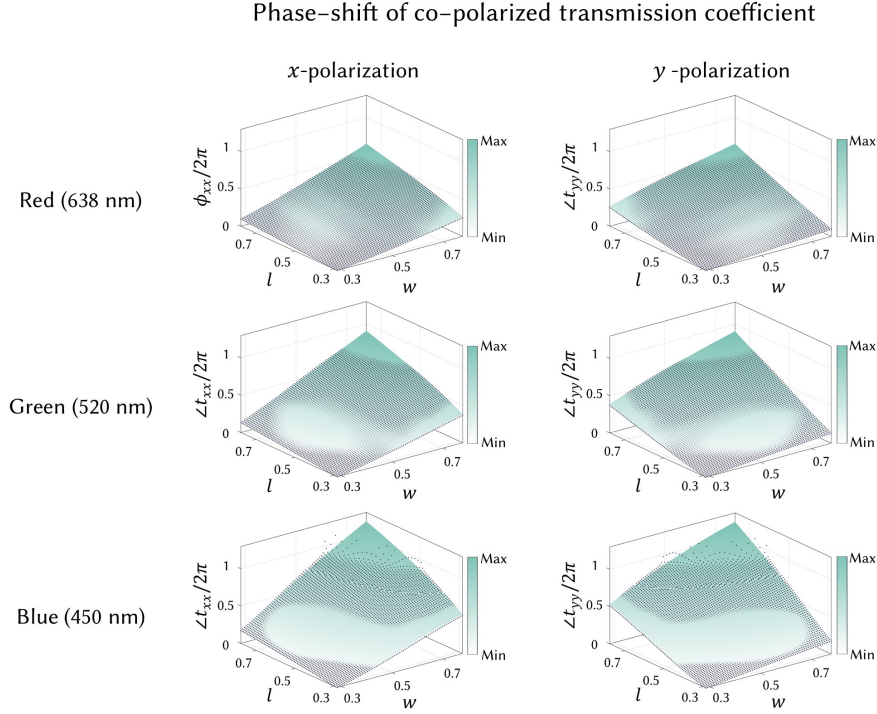

Fig. S4. Fitted surface functions and the original RCWA data. $l$ and $w$ are the normalized lengths and widths of the nanorod, in which the normalization factor is the pixel period of the metasurface. At each figure, the surface functions describe the proxy models and the simulated data are represented by the charcoal-colored point clouds.

### S1.2 Display prototype

The holographic display prototype used for experimental validation is illustrated in Figure S5. Our prototype follows the basic structure of a conventional holographic display, with a half-wave plate (HWP) and a metasurface (MS) positioned after the $4f$ system. Additionally, to facilitate metasurface alignment, an extra $4f$ system is placed after the metasurface. The light from a full-color fiber-coupled laser diode (FISBA READYBeam) is collimated using a collimating lens and directed to the 8-bit SLM (HOLOEYE LETO-3) via a beam splitter (BS). Prior to the beam splitter, a HWP and a linear polarizer (LP) are included to ensure proper polarization alignment for the SLM. The light transmitted through the SLM passes through the $4f$ system equipped with a low-pass filtering system to eliminate high-order diffraction terms. Following the first $4f$ system, an LP is positioned to filter out undiffracted terms, and an HWP on a motorized rotation mount is incorporated to control the direction of linear polarization of the light from the SLM. The metasurface is mounted on 3-axis linear stages, comprising two motorized stages in the X-axis (Thorlabs LTS300/M) and Y-axis (Thorlabs Z812B), as well as a Z-axis manual stage. These stages enable precise alignment of the SLM and the metasurface, and the motorized stages enable switching between capturing images with and without the metasurface. Finally, the metasurface plane is relayed through a second $4f$ system, and the resulting image is captured using a CCD camera (FLIR GS3-U3-51S5M-C) mounted on a motorized stage (Newport FCL100). As real images of the holograms are captured instead of virtual images with an eyepiece, the propagation distance of the hologram is calculated assuming a 50 mm eyepiece.

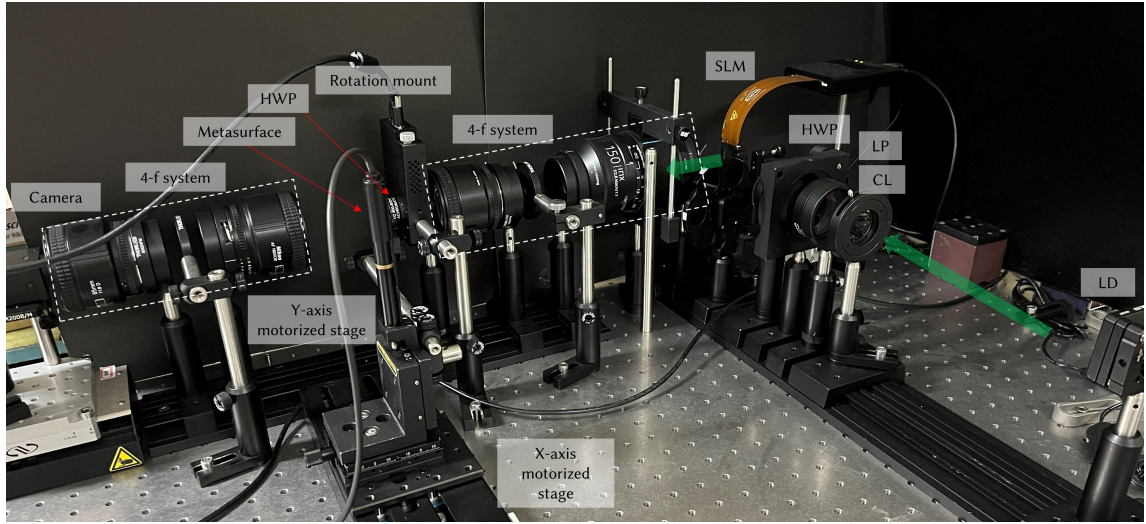

Fig. S5. Photograph or our holographic display prototype. The green arrows indicate the direction of the optical path. The optical components labeled in the photograph include laser diode (LD), SLM (spatial light modulator), collimating lens (CL), half-wave plate (HWP), linear polarizer (LP), beam splitter (BS).

### S1.3 Metasurface alignment

In Section 5.1, we discuss the utilization of the second $4f$ system in our display prototype for aligning the metasurface and the SLM. The $4f$ system allows us to directly capture the SLM plane and observe the positioning of both the SLM and the metasurface. Figure S6 shows an captured image of the relayed SLM plane, where a misalignment of $30\mu$m in both vertical and horizontal directions between the metasurface and the SLM is present. The boundary lines of the SLM and the metasurface is clearly visible, enabling manual alignment. It is worth noting that this misalignment corresponds to a shift of 10 pixels in the simulation, representing the maximum misalignment error of the noise function $f_{\text{noise}}$ employed during metasurface optimization. Since this level of misalignment is detectable by the camera, it is evident that the misalignment error in our display prototype would be much smaller than what is simulated using the noise function $f_{\text{noise}}$. Therefore, we did not conduct additional calibration steps for more precise alignment and instead relied on camera-in-the-loop training for fine-tuning.

## S2 DETAILS ON CAMERA-IN-THE-LOOP TRAINING

### S2.1 Propagation model

We use camera-in-the-loop (CITL) calibrated wave propagation model during CGH optimization for the experiments[Peng et al., 2020]. Our goal is to clarify the effect of the polarization-multiplexing metasurface, which is optimized in ideal simulation. Therefore, quality degradation from discrepancy between the simulation and the real-world system may weaken the effect of the metasurface in the experiment.

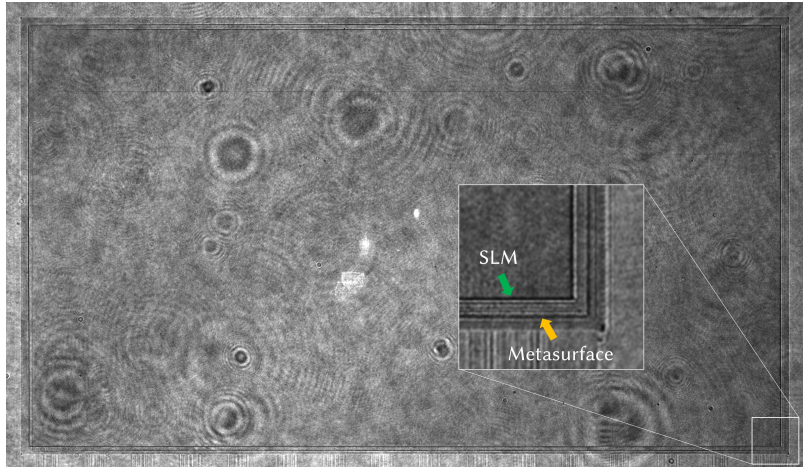

Fig. S6. The captured image shows the relayed SLM plane. In this image, a misalignment of 30 $\mu$m is present between the metasurface and the SLM in both the horizontal and vertical directions. The enlarged inset highlights that the boundary lines of the SLM and the metasurface are clearly visible.

We combine the CNNpropCNN model proposed by Choi et al. [2022] and the all-physically interpretable model by Jang et al. [2022] in our approach. Since our model aims to accurately simulate the polarization-multiplexing phenomenon, we exclude black-box models such as CNN before the metasurface. Instead, we model the nonlinear phase response of the SLM using a multi-layer perceptron [Peng et al., 2020] and incorporate the SLM pixel crosstalk noise by convolving a 3×3 kernel with the SLM phase pattern [Jang et al., 2022]. After the SLM phase mapping through the MLP and the crosstalk kernel, we apply the complex field of the light source $a_{\text{src}}, \phi_{\text{src}}$ and the metasurface, while taking into account the rotation angle of the half-wave plate. To account for potential misalignment between the fast axis of the HWP and the metasurface, we parameterize the rotation angle error of the HWP as $\theta_{\text{tilt}}$. Therefore, the Jones matrix of the HWP becomes

$$J_{\text{hwp}}(\theta; \theta_{\text{tilt}}) = \begin{bmatrix} \cos(2(\theta + \theta_{\text{tilt}})) & \sin(2(\theta + \theta_{\text{tilt}})) \\ \sin(2(\theta + \theta_{\text{tilt}})) & -\cos(2(\theta + \theta_{\text{tilt}})) \end{bmatrix}, \quad (S2)$$

where $\theta$ is the angle of the HWP for polarization rotation, with 0°, 45°, and 22.5° corresponding to horizontal, vertical, and diagonal linear polarization, respectively. For simplicity, we omit $\theta$ from Equation 9 in the manuscript.

The light from the metasurface is propagated using the modeled angular spectrum method (ASM) with a parameterized Fourier plane to account for the IRIS placed inside the $4f$ system and optical aberration. The phase aberration of the plane is modeled using Zernike polynomials up to the 9th order. After the parameterized ASM, the reconstructed amplitude passes through the CNN for image adjustment. Overall, our propagation model can be expressed as follows:

$$f_{\text{model}}(\phi) = \text{CNN}_{\text{target}}\left(f_{\text{ASM}}\left(J_{\text{proxy}}(l, w) \cdot J_{\text{hwp}}(\theta; \theta_{\text{tilt}}) \cdot a_{\text{src}} e^{i\phi_{\text{src}}} e^{i(k*\text{MLP}(\phi))}; a_{\mathcal{F}}, \phi_{\mathcal{F}}\right)\right). \quad (S3)$$

Since Jones matrices of the HWP and the metasurface have polarization-depedent elements, we capture the dataset with polarization diversity by changing the rotation angle $\theta$ of HWP. Therefore we capture the dataset with 4 different settings: without a metasurface, with the metasurface and 0° HWP, with metasurface and 22.5° HWP, and with metausrface and 45° HWP. We train our model with dataset captured with 2,000 SLM phase patterns generated from stochastic gradien descent method and the alternating direction method of multipliers method [Choi et al., 2021]. We use 5 layers U-Net for CNN$_{\text{target}}$ and optimize for 10 epochs with a learning rate of $5e^{-4}$.

### S2.2 Optimized model parameters

Figure S7 visualizes the trained physical parameters of our CITL-calibrated model, including the source intensity $a_{\text{src}}$, source phase $\phi_{\text{src}}$, amplitude $a_\mathcal{F}$ and phase $\phi_\mathcal{F}$ of the Fourier plane, SLM phase mapping through MLP, and SLM pixel crosstalk kernel. Though the phase of the Fourier plane $\phi_\mathcal{F}$ is modeled in a depthwise manner, only the phase of the central plane is showcased in the figure as a representative. Additionally, Figure S8 visualizes the trained polarization-dependent transmission coefficients of the metasurface. The model successfully captures misalignment due to shifts or distortions, as well as additional noise from dust and scratches, along with the fabricated phase patterns. The misaligned angles of the HWP are $-2.84°$, $-2.00°$, and $-1.78°$ for the red, green, and blue channels, respectively. We utilize the CITL-calibrated model for CGH optimization during the experimental validation.

## S3 ADDITIONAL RESULTS

### S3.1 Metasurface optimization result

Figure S9 visualizes the geometric parameters of the metasurface nanostructure. The left figure illustrates the schematic diagram of the metasurface nanostructures. During the metasurface optimization, the height $H$ and pixel pitch $P$ are fixed at 800 nm and 283 nm, respectively, while only the geometry maps of length $L$ and width $W$ are optimized.

The first column displays the geometry-maps of a random metasurface utilized in the simulations presented in Figure 5 and Figure 6. The geometry-maps of the random metasurface follow a uniform random distribution. The second column showcases a metasurface optimized without the noise function, which is utilized for the simulation in Figure 4. The last column illustrates the optimized metasurface with the noise function, which is actually fabricated for the experiment. The optimized metasurfaces exhibit coarser geometry-map patterns compared to the random metasurface. However, the metasurface without the noise function displays grainy, randomized patterns that make it more vulnerable to misalignment.

The power spectrum of the optimized metasurface can be found in Fig. S10. The power spectrum is derived from the Fourier transform of the complex amplitude of the metasurface. For more clear visualization, we illustrate the power spectrum is displayed on a normalized logarithmic scale. The power spectral distribution is predominantly focused on the DC component, similar to a diffuser with a narrow diffusing angle. This aligns with the interpretation of the metasurface in the manuscript Section 4, which concludes that the metasurface is optimized to have a tailored randomness.

### S3.2 Additional simulation results with partially coherent light sources

Figure S11 showcases simulation results with multiple levels of coherence. Consistent with the simulation in the manuscript, the focal length of the collimating lens is fixed to 200 mm, while we adjust the bandwidth and the aperture width of the light source. We modeled the light source's wavelength spectrum as a Gaussian distribution, with wavelength diversity represented by the standard deviation, $\sigma$. During the simulation, we first optimized the SLM phase pattern for a 2D target image using a coherent light source, and reconstructed this phase pattern with variations in the light source. The results show that the image gets blurry as the aperture size and the bandwidth increase, illustrating trade-offs in partially coherent light sources. We note that increased wavelength diversity introduces speckle noise in the image. This is because, while the speckle noise seems absent for the optimized condition in simulation, it reemerges when the reconstruction condition is different from the optimized one. However, in practice, the speckle noise is also inherent in a coherent light source, and increasing wavelength diversity reduces speckle noise at the expense of the image contrast.

### S3.3 Additional simulation and experimental results of depolarized holography

We provide additional simulation results in Figure S12 and experimentally captured results in Figure S13. Both results shows holograms with focal stack supervision. The first column represents the hologram reconstructed without the metasurface, which is equivalent to the conventional holographic displays. Second column shows the case where the metasurface inserted to the display, but only hologram with a single polarization state is captured. Third column is the depolarized holographiy, where two holograms with orthogonal polarization states are superimposed together as an intensity sum, achieving the best image quality among these three cases. An interesting observation is that even a single polarizer provides better contrast, which was not observed in the simulation. This finding contributes to the optimization of the focal stack hologram CITL. Although the peak signal-to-noise ratio (PSNR) is lower due to speckle noise, the distribution remains similar to that depicted in the histogram represented in manuscript. This indirectly implies that the degree of freedom offered by the polarization channel aids optimization, not solely in speckle reduction.

## REFERENCES

Amir Arbabi, Yu Horie, Mahmood Bagheri, and Andrei Faraon. 2015. Dielectric metasurfaces for complete control of phase and polarization with subwavelength spatial resolution and high transmission. *Nature Nanotechnology* 10, 11 (2015), 937–943.

Suyeon Choi, Manu Gopakumar, Yifan Peng, Jonghyun Kim, Matthew O'Toole, and Gordon Wetzstein. 2022. Time-Multiplexed Neural Holography: A Flexible Framework for Holographic Near-Eye Displays with Fast Heavily-Quantized Spatial Light Modulators. In *ACM SIGGRAPH 2022 Conference Proceedings* (Vancouver, BC, Canada) *(SIGGRAPH '22)*. Association for Computing Machinery, New York, NY, USA, Article 32, 9 pages.

Suyeon Choi, Manu Gopakumar, Yifan Peng, Jonghyun Kim, and Gordon Wetzstein. 2021. Neural 3D Holography: Learning Accurate Wave Propagation Models for 3D Holographic Virtual and Augmented Reality Displays. *ACM Trans. Graph.* 40, 6, Article 240 (2021), 12 pages.

Changwon Jang, Kiseung Bang, Minseok Chae, Byoungho Lee, and Douglas Lanman. 2022. Waveguide Holography: Towards True 3D Holographic Glasses.

Changil Kim, Henning Zimmer, Yael Pritch, Alexander Sorkine-Hornung, and Markus H Gross. 2013. Scene reconstruction from high spatio-angular resolution light fields. *ACM Trans. Graph.* 32, 4 (2013), 73–1.

J. P. Balthasar Mueller, Noah A. Rubin, Robert C. Devlin, Benedikt Groever, and Federico Capasso. 2017. Metasurface Polarization Optics: Independent Phase Control of Arbitrary Orthogonal States of Polarization. *Physical Review Letters* 118, 11 (2017), 113901.

Yifan Peng, Suyeon Choi, Nitish Padmanaban, and Gordon Wetzstein. 2020. Neural Holography with Camera-in-the-Loop Training. *ACM Trans. Graph.* 39, 6, Article 185 (nov 2020), 14 pages.

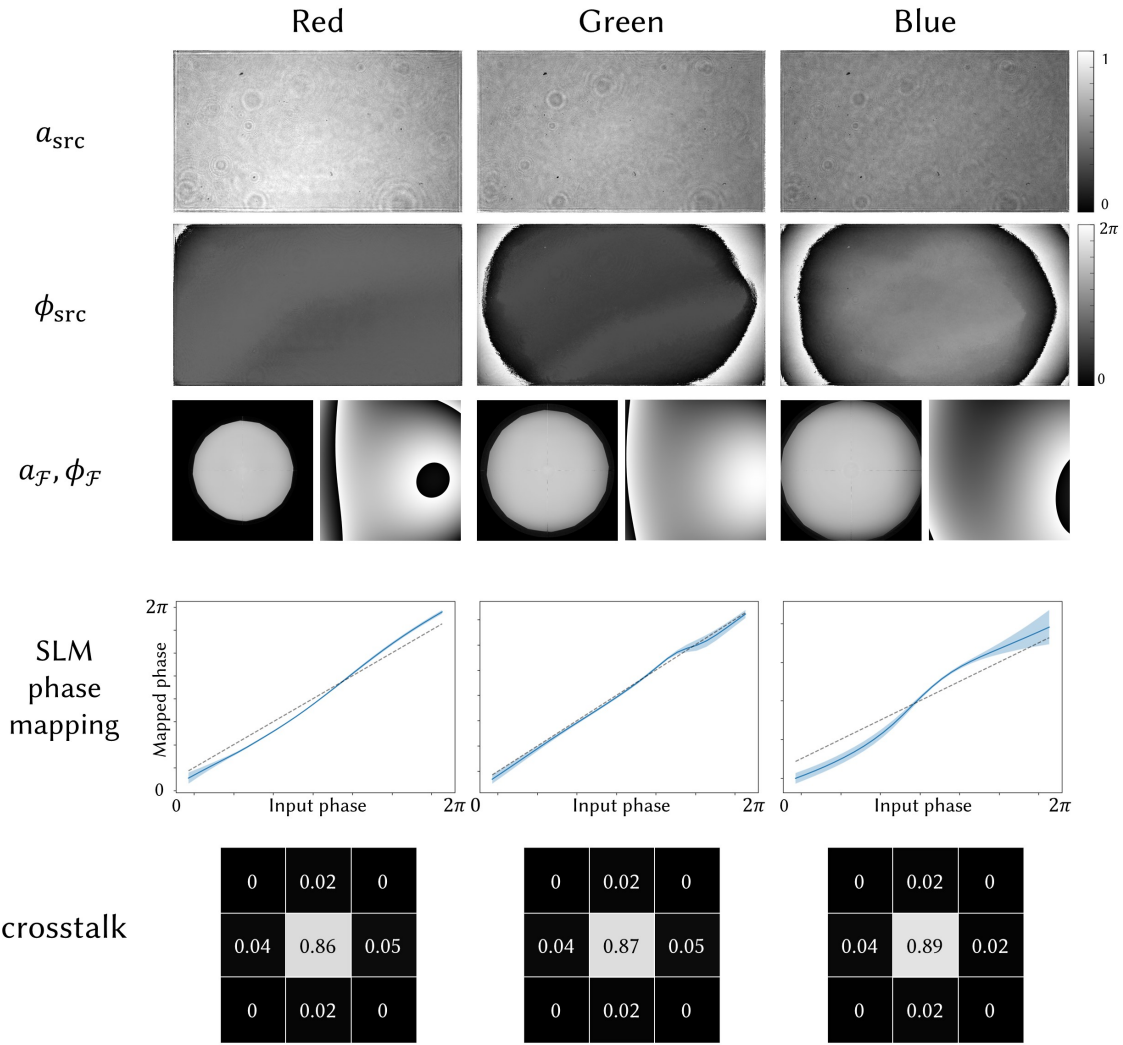

Fig. S7. Visualization of the physical parameters in the CITL-calibrated model, excluding the metasurface and HWP. In the SLM phase mapping, the dashed lines indicate the ideal mapping, while the blue solid line represents the average of the mapped phase values. The blue shaded region indicates the standard deviation. Additionally, the 3x3 crosstalk kernels are depicted in an enlarged format, with the numbers in each pixel representing the weight of the kernel.

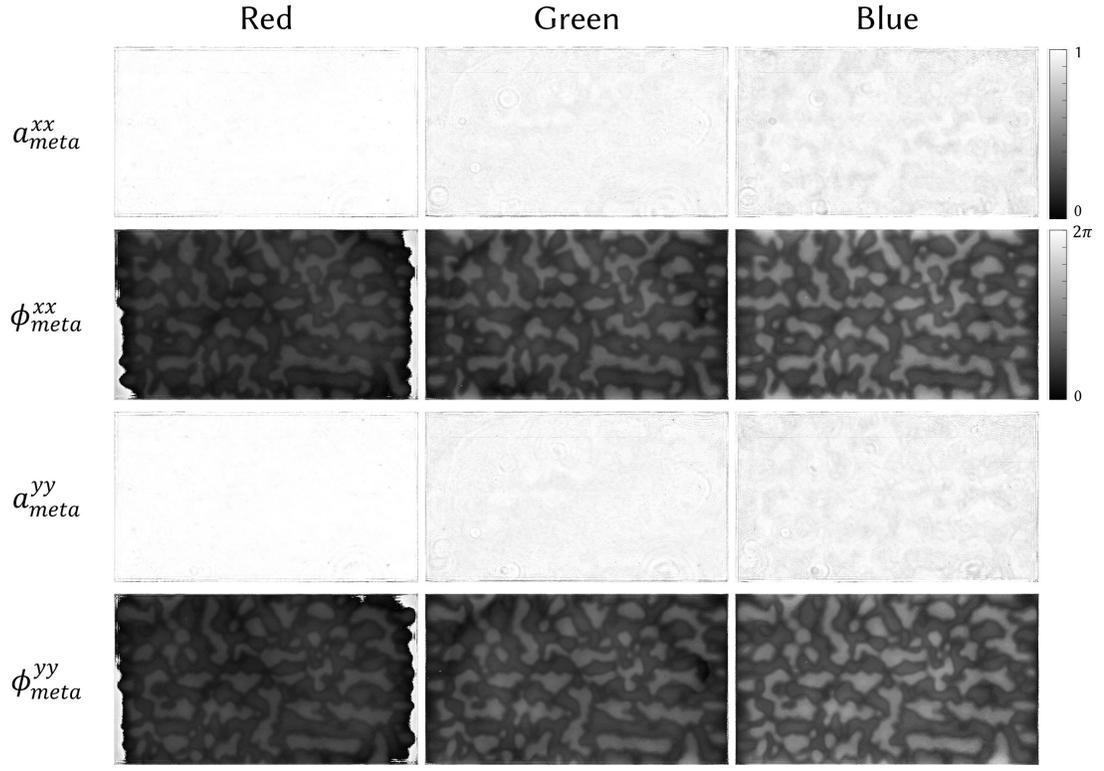

Fig. S8. Visualization of the trained amplitude and phase of the metasurface in the CITL-calibrated model. Both the phase and amplitude patterns are consistent with the ideal ones derived from the geometry-maps shown in Figure S9. The trained metasurface also includes the effects of defects from dust and scratches, as well as phase fluctuations due to the glass substrate.

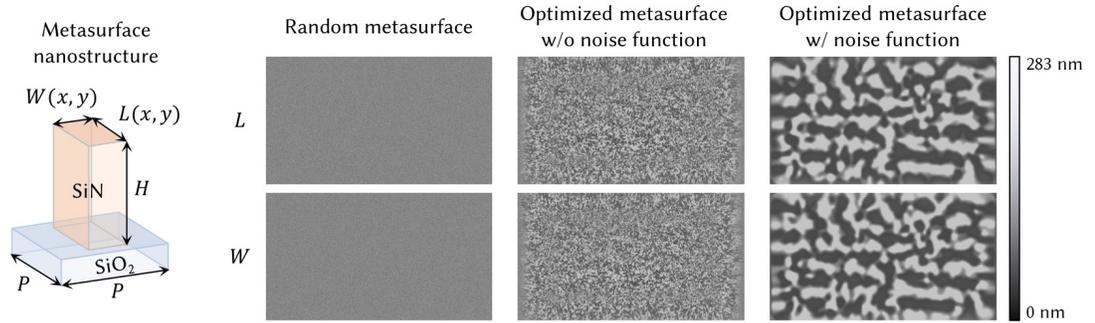

Fig. S9. (left) The schematic diagram of the metasurface nanostructure. The pixel pitch $P$, height $H$, length $L$, and width $W$ determine the transmittance of the metasurface. (right) The geometry-maps of the metasurfaces used in simulations and experiments. The dimensions of length $L$ and width $W$ are normalized with respect to the 283 nm pixel pitch

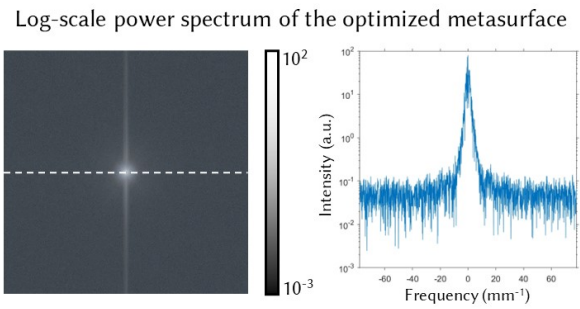

Fig. S10. Power spectral distribution of the optimized metasurface. For clarity in visualization, the power spectrum is displayed in a logarithmic scale and normalized. The plot on the right shows the cross-section of green channel from the 2D power spectrum, indicated by the white dashed line.

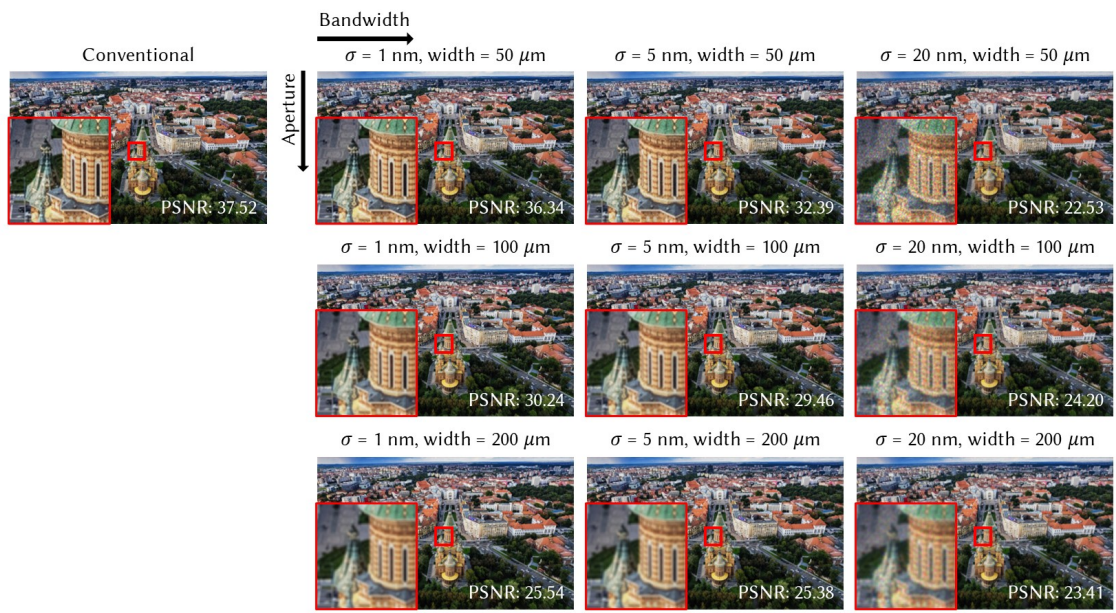

Fig. S11. Simulation results with multiple levels of coherence. The same SLM phase pattern, optimized for a coherent light source, is applied for all images simulated with different light sources. The results shows trade-offs between image contrast and speckle reduction in partially coherent light sources. Source image credits to Salomia Oana Irina.

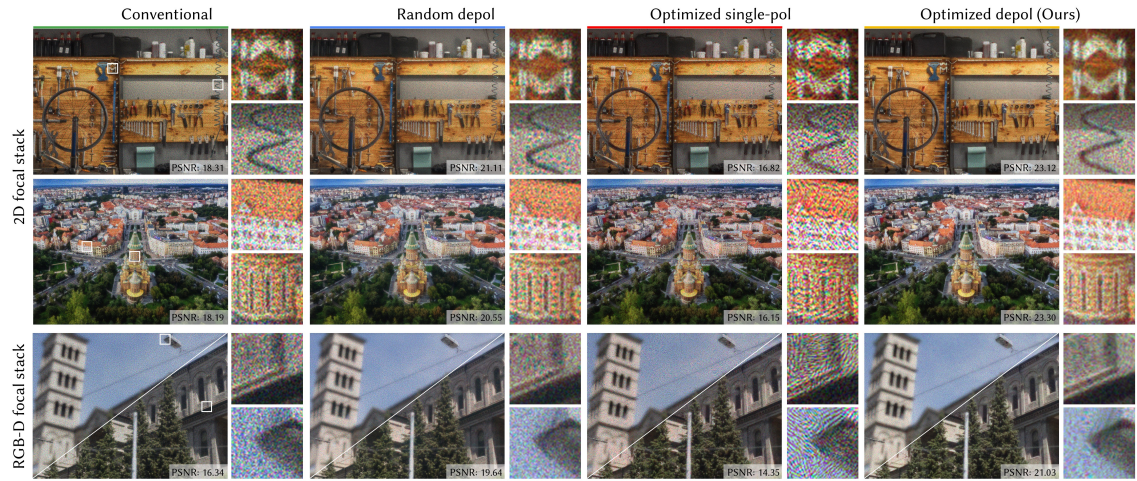

Fig. S12. Simulation results of holograms with focal stack supervision. PSNR values are reported at the bottom right corner of the image. Source images credit to BAZA Production (first row), Salomia Oana Irina (second row), and Kim et al. [2013] (third row).

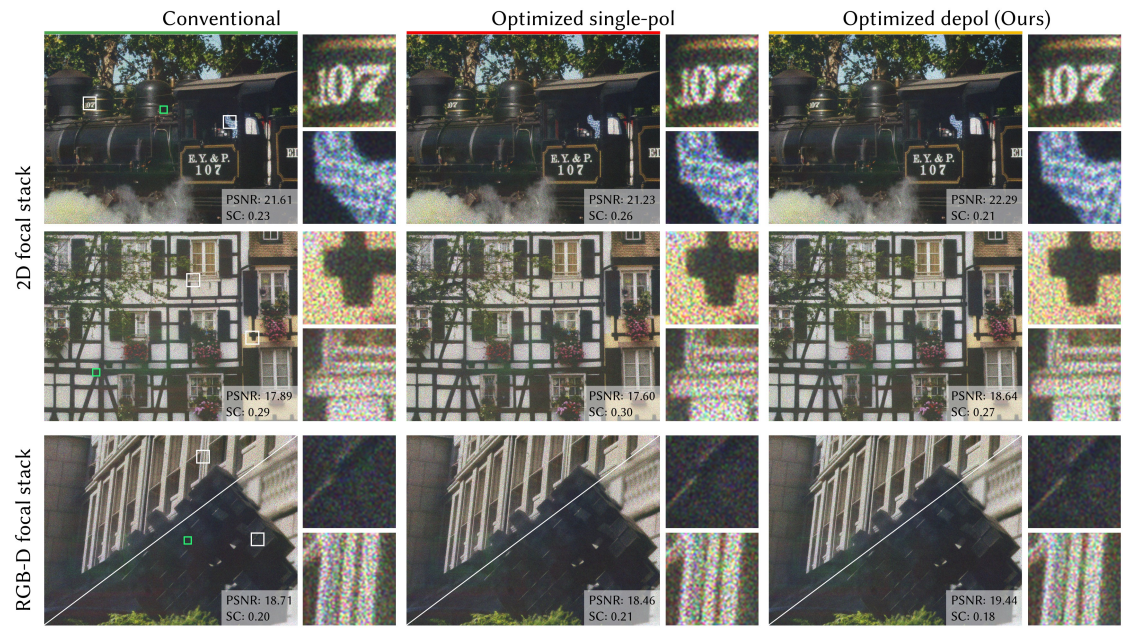

Fig. S13. Experimentally captured images of holograms with focal stack supervision. PSNR and speckle contrast values are reported at the bottom right corner of the image. The green box specifies the area that the speckle constrast is calculated. Source images credit to Bruce Raynor (first row), Pack-Shot (second row), and Kim et al. [2013] (third row).